\documentclass[amsmath,amssymb,11pt,superscriptaddress,reprint, preprintnumbers, notitlepage,aps,twocolumn,nofootinbib]{revtex4-1}
\pdfoutput=1 
\usepackage[utf8]{inputenc}
\usepackage[english]{babel}
\usepackage{amsmath}
\usepackage{graphicx}
\usepackage{dcolumn}
\usepackage{pbox}
\usepackage{amssymb}
\usepackage{epsfig}
\usepackage[dvipsnames]{xcolor}
\usepackage[normalem]{ulem}
\usepackage{slashed}
\usepackage{amssymb}
\usepackage{ mathrsfs }
\usepackage{color}
\usepackage[font=small]{caption}
\usepackage[font=small]{subcaption}
\usepackage{url}
\usepackage[makeroom]{cancel}
\usepackage{tikz}
\usetikzlibrary{arrows}
\usepackage{enumitem} 
\usepackage{soul}
\usepackage{dsfont}
\allowdisplaybreaks

\def\figureautorefname~#1\null{Fig.\,#1\null}
\def\tableautorefname~#1\null{Tab.\,#1\null}
\def\equationautorefname~#1\null{Eq.\,(#1)\null}

\definecolor{MyDarkBlue}{rgb}{0.1, 0.1, 0.8} 
\definecolor{SBlue}{rgb}{0.2, 0.4, 0.7} 
\definecolor{MyLightBlue}{rgb}{0.22,0.51,0.9}
\definecolor{MyGreen}{rgb}{0.0, 0.5, 0.0}
\definecolor{BrickRed}{rgb}{0.8, 0.25, 0.33}
\RequirePackage{hyperref}
\hypersetup{colorlinks, citecolor=SBlue,linkcolor=MyDarkBlue, urlcolor=PineGreen}

\definecolor{LightCyan}{rgb}{0.88, 1, 1}
\makeatletter
\renewcommand\@makecaption[2]{%
  \par
  \vskip\abovecaptionskip
  \begingroup
  
   \small\rmfamily
    \begingroup
     \samepage
     \flushing
     \let\footnote\@footnotemark@gobble
     \@make@capt@title{#1}{#2}\par
    \endgroup
  \endgroup
  \vskip\belowcaptionskip
}
\makeatother

\DeclareUnicodeCharacter{2212}{-}
\begin{document}
\title{\vspace{1cm}\Large 
Unified Origin of Inflation, Baryon Asymmetry, and Neutrino Mass
}

\author{\bf Ajay Kaladharan}
\email[E-mail:]{kaladharan.ajay@okstate.edu}

\affiliation{Department of Physics, Oklahom State University, Stillwater, OK 74078, USA}

\author{\bf Shaikh Saad}
\email[E-mail:]{shaikh.saad@unibas.ch}

\affiliation{Department of Physics, University of Basel, Klingelbergstrasse\ 82, CH-4056 Basel, Switzerland}

\begin{abstract}
In this work, we present a unified theoretical framework that simultaneously addresses some of the most intriguing puzzles in particle physics and cosmology, namely the origins of neutrino mass, baryon asymmetry, and cosmic inflation. 
In our model, inflation is driven by a combination of the Standard Model Higgs, the type II seesaw Higgs responsible for neutrino mass generation, and the unified symmetry-breaking Higgs field. During inflation, non-zero values of the latter field ensure the absence of the monopole problem. The baryon asymmetry is generated through the Affleck-Dine mechanism, facilitated by the non-zero angular motion in the phase of a complex scalar field, which is part of the inflaton. We find that the successful  parameter region for generating baryon asymmetry through a renormalizable term in the scalar potential requires a rather heavy type II seesaw triplet, with a mass well beyond the TeV scale. Inflationary observables, in particular, the spectral index is in excellent agree with experimental observation, whereas tensor-to scalar ratio is expected to be probed by the future LiteBIRD and CMB-S4 missions. 
\end{abstract}

\maketitle
\section{Introduction}
Although the Standard Model (SM) of particle physics has been very successful in describing the fundamental laws of physics, it fails to incorporate, for example, the observed neutrino oscillation data and the matter-antimatter asymmetry in the universe. Resolving these issues requires going beyond the SM (BSM), and Grand Unified Theories (GUTs)~\cite{Pati:1973rp,Pati:1974yy, Georgi:1974sy, Georgi:1974yf, Georgi:1974my, Fritzsch:1974nn} are the prime candidates for its ultraviolet (UV) completion. In this work, in particular, we focus on the minimal unified gauge group, namely $SU(5)$ GUT. For neutrino mass generation, within $SU(5)$ GUT, one of the simplest options  is the type II seesaw mechanism~\cite{Minkowski:1977sc,Yanagida:1979as,Glashow:1979nm,Gell-Mann:1979vob,Mohapatra:1979ia,Schechter:1980gr,Schechter:1981cv} (numerous efforts have been made to attribute non-zero masses to neutrinos in $SU(5)$ GUT. For instance, in Refs.~\cite{Dorsner:2005fq,Dorsner:2005ii,Bajc:2006ia,Dorsner:2006hw,Dorsner:2007fy,Antusch:2021yqe,Antusch:2022afk,Calibbi:2022wko,Antusch:2023kli,Antusch:2023mqe}, neutrinos acquire their masses at the tree level, while in Refs.~\cite{Wolfenstein:1980sf,Barbieri:1981yw,Perez:2016qbo,Kumericki:2017sfc,Saad:2019vjo,Dorsner:2019vgf,Dorsner:2021qwg,Antusch:2023jok,Dorsner:2024jiy,Klein:2019jgb}, the masses emerge at the loop level).

Note that the generation of neutrino mass in this setup requires only a single generation of $15_H$-dimensional Higgs representation. However, achieving a successful thermal leptogenesis demands a second copy of $15_H$ multiplet. From the point of view of minimality and predictivity, such an extension is not desirable since incorporating correct neutrino oscillation data requires only a single generation of $15_H$ multiplet. 

As recently pointed out in Ref.~\cite{Barrie:2021mwi}, one copy of scalar weak triplet with a hypercharge equal to one--employed for type II seesaw--is sufficient to give rise to the correct baryon asymmetry utilizing Affleck-Dine mechanism~\cite{Affleck:1984fy}. The prerequisites in realizing this mechanism  are (i) that the triplet carries a charge under a global $U_X(1)$ symmetry (e.g., $X=B$ (baryon number) or $L$ (lepton number)), (ii) the presence of a small term in the Lagrangian that breaks this global symmetry, and (iii) the triplet acquiring a displaced vacuum value in the early universe.

Interestingly, implementation of type II seesaw to give neutrinos non-zero masses  guarantees that the triplet scalar carries a $U_X(1)$ charge (typically the lepton number) due to its interactions with the SM fermions and the SM Higgs doublet. Moreover, certain mixed terms in the scalar potential breaks this $U_X(1)$ symmetry, which is essential to generate Majorana neutrino masses. Remarkably, this same triplet scalar can also play a role in realizing cosmic inflation~\cite{Guth:1980zm, Albrecht:1982wi, Linde:1981mu, Linde:1983gd,Barrie:2022cub}, as a result, it can naturally get a large displaced vacuum value during inflationary phase~\cite{Brout:1977ix,Sato:1981qmu,Guth:1980zm,Linde:1981mu,Albrecht:1982mp}.   Note that cosmic inflation, which refers to the rapid exponential expansion of the universe, is  another major puzzle that remains unresolved within the SM of cosmology. The theory of inflation elegantly explains several observed features of the universe, such as the large-scale homogeneity and isotropy. GUT models also require inflation to washout superheavy monopoles and other topological defects generated during the subsequent symmetry breaking phases~\cite{Kibble:1976sj,Preskill:1979zi}. However,  the exact mechanism driving this inflationary phase, as well as its connection to particle physics, remains one of the most significant open questions in cosmology.

In this work, we consider a minimal $SU(5)$ GUT in which, the  inflationary dynamics is governed by three fields:  the GUT-breaking adjoint Higgs $24_H$, the multiplet $15_H$  responsible for neutrino mass generation, and the SM Higgs contained in the fundamental representation $5_H$. Specifically, we explore inflation with a non-minimal coupling to gravity (in the context of GUT, for earlier works, see, for example, Refs.~\cite{Leontaris:2016jty,Mohapatra:2021aig,Mohapatra:2022ngo}).  Although the global $B-L$ remains preserved in the original Georgi-Glashow~\cite{Georgi:1974sy} model of $SU(5)$, in our framework, the generation of neutrino mass breaks this symmetry. This explicit breaking is controlled by a small term in the scalar potential, and the weak triplet within $15_H$, being part of the inflaton, attains a displaced vacuum value very early in the universe, ensuring the genesis of baryon asymmetry through the Affleck-Dine mechanism. Intriguingly, the adjoint field that breaks the GUT symmetry also acquires non-zero field values during inflation, thereby evading the monopole problem—a major challenge in GUT models—without the need to introduce any additional fields.  Moreover, inflationary observables are found to be in great agreement with experiments.  Our proposed model, therefore, suggests a common origin for the generation of neutrino mass, matter-antimatter asymmetry, and cosmic inflation within an economical unified theoretical framework.

The key results of this work are briefly summarized here.  In our setup, inflation is driven by three fields, whereas Ref.~\cite{Barrie:2021mwi} considered two field scenario. Additionally, due to a unified approach, we have a few significant differences compared to the earlier works. First, unlike Ref.~\cite{Barrie:2021mwi}, our model does not rely on higher-dimensional operators for generating the baryon asymmetry. Rather, a dimension four-term in the scalar potential (with coefficient $\lambda$), which explicitly breaks a global symmetry, plays a major role in producing the matter-antimatter asymmetry.  Second, as a consequence of the renormalizable term, we find a rather narrow range for the mass of the type II seesaw triplet field, specifically, in the window $10^{7}\;\mathrm{GeV}\lesssim m_\Delta \lesssim 10^{10}$ GeV (depending on the value of $\lambda$), which is consistent with the observed baryon asymmetry in the universe.  Lastly, as mentioned earlier, since the GUT-breaking Higgs participates in inflation and acquires non-zero field values during this period, our model intriguingly offers a novel resolution to the monopole problem.

This article is organized in the following way. In Sec.~\ref{sec:model}, we introduce the propose model. Fermion mass generation and gauge coupling unification are discussed in Sec.~\ref{sec:mass}.  In Sec.~\ref{sec:inflation}, the inflationary dynamics, along with the generation of baryon asymmetry, are studied. In this same section, by considering relevant washout and perturbativity constraints, we show the parameter space where the correct order of baryon asymmetry can be produced. Finally, we conclude  in Sec.~\ref{sec:con}.

\section{Model}\label{sec:model}
\textbf{Proposal}:-- In this work, we consider a GUT model based on $SU(5)$ gauge group. In this setup, the SM fermions are embedded within the $\overline 5_F$ and $10_F$ dimensional representations
\begin{align}
&\overline{5}_{F}=\begin{pmatrix}
d^C_1,\;d^C_2,\;d^C_3,\;e,\; -\nu_e
\end{pmatrix}^T,
\\
&10_{F}=\frac{1}{\sqrt{2}} \begin{pmatrix}
0&u^C_3&-u^C_2&u_1&d_1\\
-u^C_3&0&u^C_1&u_2&d_2\\
u^C_2&-u^C_1&0&u_3&d_3\\
-u_1&-u_2&-u_3&0&e^C\\
-d_1&-d_2&-d_3&-e^C&0
\end{pmatrix}.
\end{align}
In the above, we have suppressed the family index, and mass generation of the charged fermion masses will be discussed later in the text. 

Moreover, GUT symmetry is broken via the vacuum expectation value (VEV) of the adjoint $24_H$ representations. The SM Higgs doublet, which is embedded within a fundamental representation $5_H$, spontaneously breaks the electroweak (EW) symmetry. The decomposition of these Higgs fields, along with their VEV structures, are as follows:
\begin{align}
   \Phi\equiv  24_H&=\phi_8(8,1,0)+\phi_1(1,3,0)+\phi_0(1,1,0)\nonumber\\&  +\phi_3(3,2,-5/6)+\phi_{\overline{3}}(\overline{3},2,5/6),\\
  \phi\equiv  5_H&=H(1,2,1/2)+T(3,1,-1/3).
\end{align}
and,
\begin{align}
&
\langle 24_H \rangle =v_{24} \mathrm{diag}\left( -1, -1, -1, 3/2, 3/2 \right), \label{eq:24vev}
\\& \langle 5_H \rangle = (0 \quad 0 \quad 0 \quad 0 \quad v_{5}/\sqrt{2})^T \label{eq:5vev},
\end{align}
with $v_5=246$ GeV.

As a result of the GUT symmetry breaking, the superheavy vector bosons receive the following masses: \begin{align}
M_{X,Y}=\sqrt{\frac{25}{8}} g_\mathrm{GUT} v_{24},
\end{align}
where, $g_\mathrm{GUT}$ is the unified gauge coupling constant.  These gauge fields induce $d=6$ operators that are responsible for mediating proton decay. Accordingly, the lifetime of the protons can be estimated as~\cite{Langacker:1980js,Dev:2022jbf} 
\begin{align}
\tau_p\sim \frac{16\pi^2 M^4_X}{g^4_\mathrm{GUT}m^5_p},    
\end{align}
where $m_p$ is the proton mass. Then, from the current proton decay bound of $\tau_p (p\to e^+\pi^0)> 2.4\times 10^{34}$ yrs~\cite{Super-Kamiokande:2020wjk}, we obtain a lower bound on the GUT scale $M_X\sim M_\mathrm{GUT}\gtrsim 6\times 10^{15}$ GeV, where we have used $g_\mathrm{GUT}=0.6$.

Note that with the particle contents given above, neutrinos remain massless. In this work, we adopt one of the simplest ways to give neutrinos a non-zero masses, namely via type-II seesaw~\cite{Magg:1980ut,Schechter:1980gr,Lazarides:1980nt,Mohapatra:1980yp}. This requires the addition of a $15_H$ dimensional Higgs representation, which has the following decomposition: 
\begin{align}
 \Delta\equiv   &15_{H}=\Delta_{1}(1,3,1)+\Delta_{3}(3,2,1/6)+\Delta_{6}(\overline{6},1,-2/3).
\end{align}
Once the EW symmetry is broken, the weak triplet, $\Delta_{1}$, receives an induced VEV and generates neutrino mass (as will be discussed later in the text).

\textbf{Scalar potential}:--  With the above-mentioned Higgs fields, the complete  renormalizable scalar potential takes the form
\begin{align}
V&=
-m^2_\Phi Tr[\Phi^2] + \lambda^\Phi_1 Tr[\Phi^2]^2 + \lambda^\Phi_2 Tr[\Phi^4] + \mu_\Phi Tr[\Phi^3] 
\nonumber\\
&-m^2_\phi \phi^\dagger\phi + \lambda^\phi_1 (\phi^\dagger\phi)^2
\nonumber\\
&+m^2_\Delta \Delta^\dagger\Delta + \lambda^\Delta_1 (\Delta^\dagger\Delta)^2 + \lambda^\Delta_2 (\Delta^\dagger\Delta\Delta^\dagger\Delta)
\nonumber\\
&+\mu_1 \phi^\dagger\Phi\phi + \lambda^{\Phi\phi}_1 (\phi^\dagger\phi) Tr[\Phi^2]+ \lambda^{\Phi\phi}_2 (\phi^\dagger\phi \Phi \Phi) 
\nonumber\\
&+\mu_2 \Delta^\dagger\Phi\Delta 
+ \lambda^{\Phi\Delta}_1 (\Delta^\dagger\Delta) Tr[\Phi^2]+ \lambda^{\Phi\Delta}_2 (\Delta^\dagger\Delta \Phi \Phi)
\nonumber\\&
+ \lambda^{\Phi\Delta}_3 (\Delta^\dagger\Delta \Phi \Phi)^\prime 
+\lambda^{\phi\Delta}_1 (\phi^\dagger\phi)(\Delta^\dagger\Delta) +\lambda^{\phi\Delta}_2 (\phi^\dagger\phi \Delta^\dagger\Delta)
\nonumber\\&
+\bigg\{ \mu_3 \phi\phi\Delta^\dagger +\lambda_1 \phi\phi\Delta^\dagger\Phi
+h.c.\bigg\}. \label{eq:original-potential}
\end{align}
To avoid cluttering, we have suppressed all group indices. The prime here denotes  a different contraction compared to the unprimed one. The SM Higgs quartic coupling is identified as $\lambda^\phi_1=m^2_h/(2v_5^2)$.

Since $v_{24}\gg v_5\gg v_\Delta$, for all practical purposes, for the analysis of GUT symmetry breaking and computing the mass spectrum, it is sufficient to consider only non-zero $v_{24}$. The potential minimum where $SU(5)$ is broken to the SM gauge group has the stationary condition,
\begin{align}
m^2_\Phi=\left(15\lambda^\Phi_1 +\frac{7}{2}\lambda^\Phi_2 \right)v^2_{24} +\frac{3}{4}\mu_\Phi v_{24}.   
\end{align}

\textbf{Scalar mass spectrum}:--
Utilizing the above relation, the physical masses of states arising from $5_H$ are given by 
\begin{align}
m^2_H&=-m^2_\phi+\frac{3}{2}\mu_1v_{24}+\left(\frac{15}{2}\lambda^{\Phi\phi}_1+\frac{9}{4}\lambda^{\Phi\phi}_2\right)v_{24}^2,   
\\
m^2_T&=-m^2_\phi-\mu_1v_{24}+\left(\frac{15}{2}\lambda^{\Phi\phi}_1+\lambda^{\Phi\phi}_2\right)v_{24}^2. 
\end{align}
Doublet-triplet splitting is obtained by fixing 
\begin{align}
\mu_1=\frac{2}{3} \frac{m^2_\phi}{v_{24}}-\left(5  \lambda^{\Phi\phi}_1-\frac{3}{2}\lambda^{\Phi\phi}_2\right)v_{24}  .
\end{align}
The masses of the submultiplets of $24_H$ are
\begin{align}
m^2_{\phi_0}&=\frac{3}{4}\mu_\Phi v_{24}+\left(30\lambda^{\Phi}_1+7\lambda^{\Phi}_2\right)v_{24}^2,
\\
m^2_{\phi_1}&=\frac{15}{4}\mu_\Phi v_{24}+10\lambda^{\Phi}_2 v_{24}^2,
\\
m^2_{\phi_8}&=-\frac{15}{4}\mu_\Phi v_{24}+\frac{5}{2} \lambda^{\Phi}_2 v_{24}^2,
\end{align}
whereas, components of $15_H$ receive the following masses:
\begin{align}
m^2_{\Delta_1}&=m^2_\Delta+ \frac{3}{2}\mu_2 v_{24}+\left(
\frac{15}{2} \lambda^{\Phi\Delta}_1
+\frac{9}{4} \lambda^{\Phi\Delta}_2
+\frac{9}{4} \lambda^{\Phi\Delta}_3
\right)v_{24}^2,
\\ 
m^2_{\Delta_3}&=m^2_\Delta+ \frac{1}{4}\mu_2 v_{24}+\left(
\frac{15}{2} \lambda^{\Phi\Delta}_1
+\frac{13}{8} \lambda^{\Phi\Delta}_2
-\frac{3}{2} \lambda^{\Phi\Delta}_3
\right)v_{24}^2,
\\
m^2_{\Delta_6}&=m^2_\Delta-\mu_2 v_{24}+\left(
\frac{15}{2} \lambda^{\Phi\Delta}_1
+ \lambda^{\Phi\Delta}_2
+ \lambda^{\Phi\Delta}_3
\right)v_{24}^2.
\end{align}
In the subsequent sections, we examine the inflationary dynamics and check the consistency of the mass spectrum of the scalar fields as computed above with a successful generation of baryon asymmetry.

\textbf{Global symmetry:}
Note, in the Georgi-Glashow $SU(5)$ model, although baryon number $B$ and lepton number $L$ are separately broken,  $B-L$ is still a global symmetry; therefore, $B-L$ is conserved (hence, all nucleon decay operators conserved $B-L$). Consequently, neutrinos are massless. However, in extended scenarios, such as ours with $\Delta$, this global $B-L$ symmetry is broken. Consequently, neutrinos receive non-zero masses. For example, in the scalar potential, both the terms $\mu_3$ and $\lambda_1$ violate it (however, Yukawa couplings respect this symmetry). 
More specifically, the unbroken abelian global symmetry, $U(1)_X$, is identified as~\cite{Wilczek:1979hc} 
\begin{align}
X=5(B-L)-4Y    
\end{align}
such that $X[\overline 5_F]=-3$ and $X[10_F]=+1$. Therefore,  the scalars carry the following charges under $U(1)_X$:
\begin{align}
X[\phi]=-2,\; 
X[\Delta]=+6,\;
X[\Phi]=0.\;
\end{align} 

From these charge assignments, one can see that the cubic term $\mu_3$ and the quartic term $\lambda_1$ in the scalar potential Eq.~\eqref{eq:original-potential} explicitly break the global $U(1)_X$ symmetry.   As will be shown below, the latter term is responsible~\cite{Barrie:2024yhj} for generating the matter-antimatter asymmetry in the universe. The cubic term, on the other hand, must be very small so as not to ruin~\cite{Barrie:2022cub} the predictability of the generated asymmetry during the inflation.

\section{Fermion Mass Generation}\label{sec:mass}
\textbf{Charged Fermion Sector}:--
Among the scalar fields introduced in our model, only the fundamental Higgs can provide masses to the charged fermions. The relevant part of the Yukawa couplings reads
\begin{align}
&\mathcal{L}_Y\supset Y_{10}^{ab}10_F^a10^b_F5_{H}+Y_{5}^{ai} 10^a_F  \overline 5^i_F 5^*_{H}.
\end{align}
Once the EW symmetry is spontaneously broken, charged fermions acquire the following masses:
\begin{align}
 M_U &=\sqrt{2}v_5\left(Y^u+Y^{u\,T}\right),\label{eq:uncorrected_U}\\
M_E &= \frac{v_5}{2}Y^{d\,T},\label{eq:uncorrected_E}\\
M_D &= \frac{v_5}{2}Y^d. \label{eq:uncorrected_D}
\end{align}
Since it predicts $M_E=M_D^T$ at the GUT scale, this simplest setup fails to correctly reproduce the observed mass spectrum. Therefore, we introduce a pair of vectorlike fermions  $10_F+\overline{10}_F$ that allows additional Yukawa interactions of the form
\begin{align}
&\mathcal{L}_Y\supset  y^\prime \overline{10}_F\overline{10}_F5^*_{H} + \left(m_a  + \lambda_a 24_H\right) \overline{10}_F  10_F^a,
\end{align}
where $a=1-4$. With these interactions, the modified mass matrices of the down-type quarks and the charged leptons are given by~\cite{Antusch:2023mqe}
\begin{align}
&M_D=
\begin{pmatrix}
\left(Y_{5}\right)^{ij}\frac{v_5}{2}
&
m_i-\frac{\lambda_i v_{24}}{4} 
\\
\left(Y_{5}\right)^{4j}\frac{v_5}{2}
&
m_4-\frac{\lambda_4 v_{24}}{4} 
\end{pmatrix}_{4\times 4},
\\
&M_E=
\begin{pmatrix}
\left(Y_{5}^T\right)^{ij}\frac{v_5}{2}
&
\left(Y_{5}^T\right)^{i4}\frac{v_5}{2}
\\
m_j- \frac{3}{2}\lambda_j v_{24}  
&
m_4-\frac{3}{2}\lambda_4 v_{24} 
\end{pmatrix}_{4\times 4}.
\end{align}
Owing to the mixing with the vector-like fermions, the above mass matrices break the wrong mass relation, namely $M_E=M_D^T$, and a consistent fit to the charged fermion spectrum can be easily obtained.

\textbf{Gauge coupling unification:}
In the $SU(5)$ setup, gauge coupling unification requires a few states to live below the GUT scale. Interestingly, within our scenario, a pair of vectorlike fermions employed to cure the wrong mass relations can greatly help in achieving coupling unification.  As shown in Ref.~\cite{Antusch:2023mqe}, the vectorlike quarks $\widetilde Q(3,2,1/6)\subset 10_F$ having a mass in the range $m_{\widetilde Q}\sim 10^3-10^6$ GeV can provide gauge coupling unification consistent with current proton decay limits from Super-Kamiokande. In this analysis, the weak triplet $\phi_1$ and the color octet $\phi_8$ from $24_H$ are assumed to live in the intermediate mass scale in between $m_{\widetilde Q}-M_\mathrm{GUT}$.

\textbf{Neutrino Mass}:--
In our framework, neutrino mass is generated by the type II seesaw. The corresponding Yukawa interaction is given by
\begin{align}
&\mathcal{L}_Y\supset \frac{1}{2} Y^{ij}_\nu \overline 5_F^i \overline  5_F^j 15_H
\\& \supset
\frac{1}{2} \chi_2 e^{i\theta_2}  \bigg\{\frac{1}{\sqrt{2}}
\left(Y_\nu\right)_{ij} \bigg\} \nu^T_i C \nu_j.
\end{align}
From the above Lagrangian, neutrinos receive the following masses:
\begin{align}
m^\nu_{ij}= v_\Delta    \left(Y_\nu\right)_{ij} ,
\end{align}
where the induced VEV of the neutral component of the  $15_H$ multiplet has the expression
\begin{align}
v_\Delta&= -\frac{v^2_5}{2m^2_{\Delta_1}} \underbrace{ \left( \mu_3+ \frac{3}{\sqrt{2}}\lambda_1 v_{24} \right) }_{\equiv \mu}. \label{nu:constraint}
\end{align}
Recall, $v_5=246$ GeV and $v_{24}\sim 10^{16}$ GeV; therefore, if the second term dominates, the size of $\lambda_1$ must be somewhat small to replicate the neutrino mass scale.  As discussed above, to predict the generated baryon asymmetry at the end of the inflation, the cubic term is taken to be rather small. Therefore, within our setup, we have an one-to-one correspondence between $\lambda_1$ and $v_\Delta$.   Although these parameters $\mu_3$ and $\lambda_1$ do not enter in the expressions of the scalar masses, they are restricted from the EW-precision  (upper bound) and neutrino mass constraints (lower bound due to perturbativity)
\begin{align}
\mathcal{O}(1)\;\mathrm{GeV}> v_\Delta \gtrsim 5\times 10^{-11}\;\mathrm{GeV}.  
\end{align}

\section{Inflationary Dynamics}\label{sec:inflation}
Cosmic inflation addresses the horizon and flatness problems of standard Big Bang cosmology and explains the origin of structure formation in the observable universe. Moreover, the unwanted superheavy  monopoles generated~\cite{Kibble:1976sj} during the phase transition
\begin{align}
SU(5) \xrightarrow[]{\langle 24_H\rangle} SU(3)_C \times SU(2)_L \times U(1)_Y  \end{align}
are also diluted away by inflation. Without inflation, these stable monopoles would overclose the universe. Our proposed scenario is particularly compelling because inflation is achieved using only the minimal set of fields introduced in the previous section.

The inflationary dynamics will be
induced by a combination of the SM Higgs $H\in 5_H$, the weak triplet $\Delta_1\in 15_H$, and the GUT breaking field $\phi_0\in 24_H$. To realize inflation, one must guarantee flat directions in the scalar potential, which we achieve through these fields having  non-minimal couplings to gravity that generically arise in curved spacetime~\cite{Chernikov:1968zm}. These couplings flatten the scalar potential at large field values, ensuring that the slow-roll parameters are adequately satisfied and that the predicted observational signatures match current CMB measurements~\cite{Planck:2018jri}.  Inflation of this type has been consider, for example, in Refs.~\cite{Bezrukov:2007ep,Bezrukov:2008ut,Garcia-Bellido:2008ycs,Barbon:2009ya,Barvinsky:2009fy,Bezrukov:2009db,Giudice:2010ka,Bezrukov:2010jz,Burgess:2010zq,Lebedev:2011aq,Lee:2018esk,Choi:2019osi,Barrie:2021mwi,Barrie:2022cub,Sopov:2022bog,Barrie:2024yhj}.

For the consideration of inflation, the only relevant fields are the neutral components that can acquire VEVs, which we denote by
\begin{align}
&(1,2,1/2)\supset \phi^0=\frac{1}{\sqrt{2}} \rho_1 e^{i\theta_1}, \label{eq:field01}\\
&(1,3,1)\supset \Delta^0=\frac{1}{\sqrt{2}}\rho_2 e^{i\theta_2},\label{eq:field02}\\
&(1,1,0)\supset \Phi^0=\rho_3.\label{eq:field03}
\end{align}
Then the Lagrangian, including the non-minimal couplings to gravity, can be written as (in the Jordan ($\mathcal{J}$) frame)
\begin{align}
\frac {\mathcal {L}^\mathcal{J}}{\sqrt{-g^\mathcal{J}}}&\supset -\left ( \frac {M^2_{\mathrm{pl}}}{2}+\xi_{\phi}\phi^{\dagger}_0\phi_0+\xi_{\Delta}\Delta^{\dagger}_0\Delta_0 +\frac{\xi_{\Phi}}{2}\Phi^{2}_0\right )\mathcal{R}^\mathcal{J} 
\nonumber\\&
-g^{\mu\nu} \left(D_\mu \phi^0\right)^\dagger\left(D_\mu \phi^0\right)
-g^{\mu\nu} \left(D_\mu \Delta^0\right)^\dagger\left(D_\mu \Delta^0\right)
\nonumber\\&
-g^{\mu\nu} \left(D_\mu \Phi^0\right)^\dagger\left(D_\mu \Phi^0\right)
-V_\mathrm{inf}
+\mathcal {L}_\mathrm{Yukawa}.
\end{align}
Here $\xi_{\phi}, \xi_{\Delta}, \xi_{\Phi}$ are real dimensionless (non-minimal) couplings which are taken to be positive, $\mathcal{R}^\mathcal{J}$ is the Ricci scalar in the Jordan
frame, and $M_{\mathrm{pl}}$ is the reduced
Planck mass. Moreover, with the fields defined in Eqs.~\eqref{eq:field01}-\eqref{eq:field03}, the scalar potential for inflation $V_\mathrm{inf}$ takes the following form:
\begin{align}
V_\mathrm{inf}=& 
-\frac{1}{2}m^2_\phi \rho_1^2 +\underbrace{ \frac{\lambda^\phi_1}{2} }_{\equiv \lambda_\phi/8} \rho_1^4
+\frac{1}{2}m^2_\Delta \rho_2^2 +\underbrace{\left( \frac{\lambda^\Delta_1}{4}+\frac{\lambda^\Delta_2}{4} \right)}_{\equiv \lambda_\Delta/8} \rho_2^4
\nonumber\\&
-\frac{1}{2} m^2_\Phi \rho_3^2 + \underbrace{\left( \frac{\lambda^\Phi_1}{4}+\frac{7\lambda^\Phi_2}{120} \right)}_{\equiv \lambda_\Phi/8 }\rho_3^4
+\underbrace{ \left( \frac{\lambda^{\phi\Delta}_1}{4}+\frac{\lambda^{\phi\Delta}_2}{4} \right) }_{\equiv \lambda_{\phi\Delta}/4}\rho_1^2\rho_2^2
\nonumber\\&
+\underbrace{ \left( \frac{\lambda^{\Phi\phi}_1}{4}+\frac{3\lambda^{\Phi\phi}_2}{40} \right) }_{\equiv \lambda_{\Phi\phi}/4}\rho_1^2\rho_3^2
\nonumber\\&
+\underbrace{  \left( \frac{\lambda^{\Phi\Delta}_1}{4}+\frac{\lambda^{3\Phi\Delta}_2}{40} +\frac{3\lambda^{\Phi\Delta}_3}{40} \right) }_{\equiv \lambda_{\Phi\Delta}/4}\rho_2^2\rho_3^2
\nonumber\\&
+  \underbrace{ \frac{\sqrt{3}\mu_1}{4\sqrt{5}} }_{\equiv \hat\mu_1} \rho_1^2\rho_3
+  \underbrace{ \frac{\sqrt{3}\mu_2}{4\sqrt{5}} }_{\equiv \hat\mu_2} \rho_2^2\rho_3
+  \underbrace{ \frac{\mu_\Phi}{4\sqrt{15}} }_{\equiv \hat\mu_4} \rho_3^3
\nonumber\\&
+\bigg\{   \underbrace{ \lambda_1 \frac{\sqrt{3}}{4\sqrt{10}} }_{\equiv \lambda} e^{i(2\theta_1-\theta_2)} \rho_1^2\rho_2\rho_3+h.c.\bigg\}
\nonumber\\&
+\bigg\{  \underbrace{ \frac{\mu_3}{2\sqrt{2}} }_{\equiv \hat\mu_3} e^{i(2\theta_1-\theta_2)} \rho_1^2\rho_2+h.c.\bigg\}. \label{Vinf}
\end{align}
In this work, we will assume the cubic couplings to be sufficiently small (that can be easily arranged) so that they do not affect the inflation dynamics.

In this setup, the inflation phase starts when one or more of the modulus fields are displaced from their minimum, entering a large-field regime, i.e.,
$\xi_{\phi}\rho_1^2+\xi_\Delta \rho_2^2+\xi_\Phi\rho_3^2 \gg M^2_\mathrm{pl}$. For the analysis of the inflationary dynamics, it is customary to work in the Einstein frame ($\mathcal{E}$), which is related to the Jordan frame  through a local rescaling of the spacetime metric:
\begin{equation}
g^{\mathcal{E}}_{\mu\nu}=\Omega^2(\rho_1,\rho_2,\rho_3)g^{\mathcal{J}}_{\mu\nu},
\end{equation}
where, $\Omega^2(\rho_1,\rho_2,\rho_3)=1+\frac {\xi_{\phi}\rho_1^2+\xi_\Delta \rho_2^2+\xi_\Phi\rho_3^2}{M_{\mathrm{pl}}^2}$.\\
This Weyl transformation serves to both restore the minimal coupling to gravity and flatten the potential for large values of the modulus fields. However, in this Einstein  frame, the metric, $g^{\mathcal{E}}_{\mu\nu}$,  is no longer trivial due to the  non-canonical form of the kinetic terms~\cite{Kaiser:2010ps}.    In the large-field regime ($\Omega^2\gg 1$), the Einstein frame scalar potential can be written as
\begin{align}
&V^\mathcal{E}(\rho_i)=\Omega^{-4}V^{\mathcal{J}}(\rho_i) \\
&\approx\frac {M_{\mathrm{pl}}^4}{8} (\xi_{\phi}\rho_1^2+\xi_\Delta \rho_2^2+\xi_\Phi\rho_3^2)^{-2} \bigg\{{\lambda_\phi} \rho_1^4+{\lambda_\Delta} \rho_2^4+{\lambda_\Phi} \rho_3^4
\nonumber\\&
+2{\lambda_{\phi\Delta}}\rho_1^2\rho_2^2
+2{\lambda_{\Phi\phi}}\rho_1^2\rho_3^2
+2{\lambda_{\Phi\Delta}}\rho_2^2\rho_3^2+16\lambda \cos(\theta_2)\rho_1^2\rho_2\rho_3\bigg\}. \label{potential0}
\end{align}
In the above potential, the only term that violates the global $U(1)_X$ symmetry is the one with coefficient $\lambda$, which needs to be very small so that it does not spoil the inflationary trajectory. This term, however, plays a  crucial role in generating baryon asymmetry.  In this three-field inflationary scenario, hyper-valley condition for realizing inflation successfully demands~\cite{Sopov:2022bog}  
\begin{equation}
A_{\phi \Delta}>0,\,\, A_{\phi \Phi}>0,\,\, A_{\Phi \Delta}>0 \;,
\end{equation}
where,
\begin{align}
          \kappa_{ij}&\equiv\lambda_{ij}\xi_i-\lambda_i\xi_j, \\
\gamma_{ij}=\gamma_{ji}&\equiv\lambda_i\lambda_j-\lambda_{ij}^2>0,\\
A_{ij}=A_{ji}&\equiv\xi_k\gamma_{ij}+\lambda_{ik}\kappa_{ji}+\lambda_{jk}\kappa_{ij}.
\end{align}
As we will show later with an example benchmark point that  all these conditions can be satisfied for $\mathcal{O}(1)$ quartic couplings. For the analysis, we use the following parametrisation for the underlying field space:
\begin{equation}
    \rho_1=\varphi \cos \alpha \sin\beta, \,\, \rho_2=\varphi \sin \alpha \sin\beta, \,\, \rho_3=\varphi\cos \beta.
\end{equation}
where, $\alpha$ and $\beta$ are defined as,
\begin{equation}
    \cos^2 \alpha=\frac {A_{\Delta \Phi}}{A_{\Delta \Phi}+A_{\phi \Phi}},\,\,\cos^2 \beta=\frac {A_{\phi \Delta}}{A_{\phi \Delta}+A_{\phi \Phi}+A_{\Delta \Phi}}.
    \label{eq:inflatangle}
\end{equation}

The above parametrization leads to 
\begin{equation}
    \Omega^2=1+\xi \frac{\varphi^2}{M_{\mathrm{pl}}^2},
\end{equation}
with
\begin{equation}
    \xi=\xi_{\phi} \cos^2\alpha \sin^2\beta+\xi_{\Delta} \sin^2 \alpha \sin^2\beta+\xi_{\Phi}\cos^2 \beta.
\end{equation}

In terms of $\varphi$, the Lagrangian in the Jordan Frame is given by,
\begin{align}
     \frac {\mathcal {L}^\mathcal{J}}{\sqrt{-g^\mathcal{J}}}&=-\frac {M_{\mathrm{pl}^2}}{2}\mathcal{R}^\mathcal{J} -\frac {\xi}{2}\varphi^2\mathcal{R}^\mathcal{J} -\frac 12 g^{\mu \nu}\partial_\mu \varphi \partial_\nu \varphi \nonumber\\
     &-\frac 12 \varphi^2 \sin^2 \alpha \sin^2 \beta  g^{\mu \nu}\partial_\mu \theta_2 \partial_\nu \theta_2-V(\varphi,\theta_2),
\end{align}
where the potential is given by 
\begin{equation}
     V(\varphi,\theta_2)=-\frac 12 m^2\varphi^2+(\tilde\mu+2\hat{\mu}\cos \theta_2)\varphi^3+\frac 14 \tilde\lambda \varphi^4+2\lambda^\prime\cos \theta_2 \varphi^4 , \label{eq:Vphi}
\end{equation}
and we have defined the following quantities:
\begin{align}
    &m^2=m^2_{\phi}\cos^2\alpha \sin^2 \beta+m^2_{\Delta}\sin^2 \alpha \sin^2 \beta+ m^2_{\Phi} \cos^2 \beta, \\
&\tilde\mu=\hat\mu_1 \cos^2\alpha \sin^2 \beta \cos \beta+\hat{\mu}_2\sin^2 \alpha \sin^2 \beta \cos{\beta}+\hat{\mu}_4\cos^3\beta,\\
&\hat\mu=\hat {\mu}_3 \cos^2 \alpha\sin \alpha \sin^3 \beta,\\
&\tilde{\lambda}=\frac {\lambda_\phi}{2}\cos^4\alpha \sin^4 \beta+\frac{\lambda_{\Delta}}{2}\sin^4 \alpha \sin^4 \beta+\frac {\lambda_{\Phi}}2 \cos^4 \beta \nonumber\\
&+\lambda_{\phi \Delta} \cos^2\alpha \sin^2\alpha \sin^4 \beta + \lambda_{\phi \Phi}\cos^2 \alpha \sin^2 \beta \cos^2 \beta\nonumber \\
&+\lambda_{\Delta \Phi}\sin^2 \alpha \sin^2 \beta \cos^2 \beta, \\
&\lambda^\prime=\lambda \cos^2 \alpha \sin \alpha \sin^3 \beta \cos \beta .
\end{align}

Now reparametrizing $\varphi$ in terms of canonically normalized field, which we denote by $\chi$,
\begin{equation}
    \frac{\mathrm{d} \chi}{\mathrm{d} \varphi}=\frac {\sqrt{6\xi^2\varphi^2/M^2_{\mathrm{pl}}+\Omega^2}}{\Omega^2},
\end{equation}
it allows us to write $\chi$ in terms of $\varphi
$,
\begin{align}
    \frac {\chi(\varphi)}{M_{\mathrm{pl}}}=\frac 1\xi&\left ( \sqrt{1+6\xi}\sinh^{-1}\left ( \sqrt{\xi+6\xi^2}\frac {\varphi}{M_{\mathrm{pl}}} \right ) \right. \nonumber\\
&\left. -\sqrt{6\xi}\sinh^{-1}\left ( \sqrt{6\xi^2}\frac {\varphi}{M_{\mathrm{pl}}}/\sqrt{1+\xi\frac {\varphi^2}{M^2_{\mathrm{pl}}}} \right )  \right ).
\end{align}
Therefore, the Lagrangian in the Einstein frame takes the following form:  
\begin{align}
    \frac {\mathcal{L}^\mathcal{E}}{\sqrt{-g}}=&-\frac {M^2_{\mathrm{pl}}}{2}R-\frac12 g^{\mu \nu}\partial_\mu \chi\partial_\nu \chi \nonumber\\
   & -\frac 12 f(\chi)g^{\mu \nu}\partial_\mu \theta_2 \partial_\nu \theta_2-U(\chi,\theta_2),
\end{align}
where 
\begin{equation}
    f(\chi)=\frac {\varphi(\chi)^2\sin^2\alpha \sin^2 \beta}{\Omega^2},\, \quad
U(\chi,\theta_2)=\frac {V(\varphi(\chi),\theta_2)}{\Omega^4}.
\end{equation}

From the above Lagrangian, the equations of motion become,
\begin{align}
    &\ddot{\chi}-\frac 12 f_{,\chi}\dot{\theta_2}^2+3H\dot{\chi}+U_{,\chi}=0,\\
&\ddot{\theta_2}+\frac {f_{,\chi}}{f(\chi)}\dot{\theta_2}\dot{\chi}+3H\dot{\theta}_2+\frac {1}{f(\chi)}U_{,\theta_2}=0,
\end{align}
and the  Hubble parameter is given by
\begin{equation}
    H^2=\frac {1}{3M^2_p}\left ( \frac 12 f(\chi)\dot{\theta}_2^2+\frac 12 \dot{\chi}^2+U(\chi,\theta_2) \right ).
\end{equation}
Using slow-roll approximation, one can straightforwardly finds,
\begin{equation}
    \dot{\chi}\approx -\frac {M_p U_{,\chi}}{\sqrt{3U}}\, ,\quad \quad
\dot{\theta}_2 \approx -\frac {M_p U_{,\theta_2}}{f(\chi)\sqrt{3U}}. \label{eq:U}
\end{equation}

We reparametrizing $\tau=t H_0$, where $H_0=m_s/2$  ($m_s=3\times 10^{13}\, \mathrm{GeV}$ is the Starobinsky mass scale~\cite{Starobinsky:1980te}) and denote a derivative with respect to $\tau$ by a prime. We, therefore, get
\begin{align}
    &\chi^{\prime \prime}-\frac 12f_{,\chi}{\theta_2^\prime}^2+3\frac {\tilde{H}}{M_{\mathrm{pl}}}\chi^\prime+\frac {U_{,\chi}}{H_0^2}=0,\\
&\theta_2^{\prime \prime}+\frac {f_{,\chi}}{f(\chi)}\theta_2^\prime\chi^\prime+3\frac {\tilde{H}}{M_{\mathrm{pl}}}\theta_2^\prime+\frac {1}{f(\chi){H}_0^2}U_{,\theta_2}=0,
\end{align}
where we have defined the reduced Hubble parameter as
\begin{equation}
    \tilde{H}^2=\frac 13\left ( \frac 12 {\chi^\prime}^2+\frac 12 f(\chi){\theta_2^\prime}^2+\frac {U}{H_0^2} \right ).
\end{equation}

\subsection{Inflationary Observables}
\begin{table*}[t]
\centering
\setlength{\tabcolsep}{8pt} 
\begin{tabular}{|c|c|c|c|c|c|c|c|c|}
\hline
$\xi_\phi$ & $\xi_\Delta$ & $\xi_\Phi$ & $\lambda_\phi$ & $\lambda_\Delta$ & $\lambda_{\Phi}$ & $\lambda_{\phi \Delta}$ & $\lambda_{\phi \Phi}$ & $\lambda_{\Delta \Phi}$ \\
\hline
$234.746517$  & $187.002557$ & $187.242948$  & $0.519288$ &
$0.636964$ & $0.467765$& $-0.474554$&$-0.323258$&$0.062880$ \\
\hline
\end{tabular}
\caption{Non-minimal couplings to gravity and quartic couplings for the benchmark point.   }
\label{tab:BP}
\end{table*}

\begin{figure}[!tb]
    \centering
\includegraphics[width=0.5\textwidth]{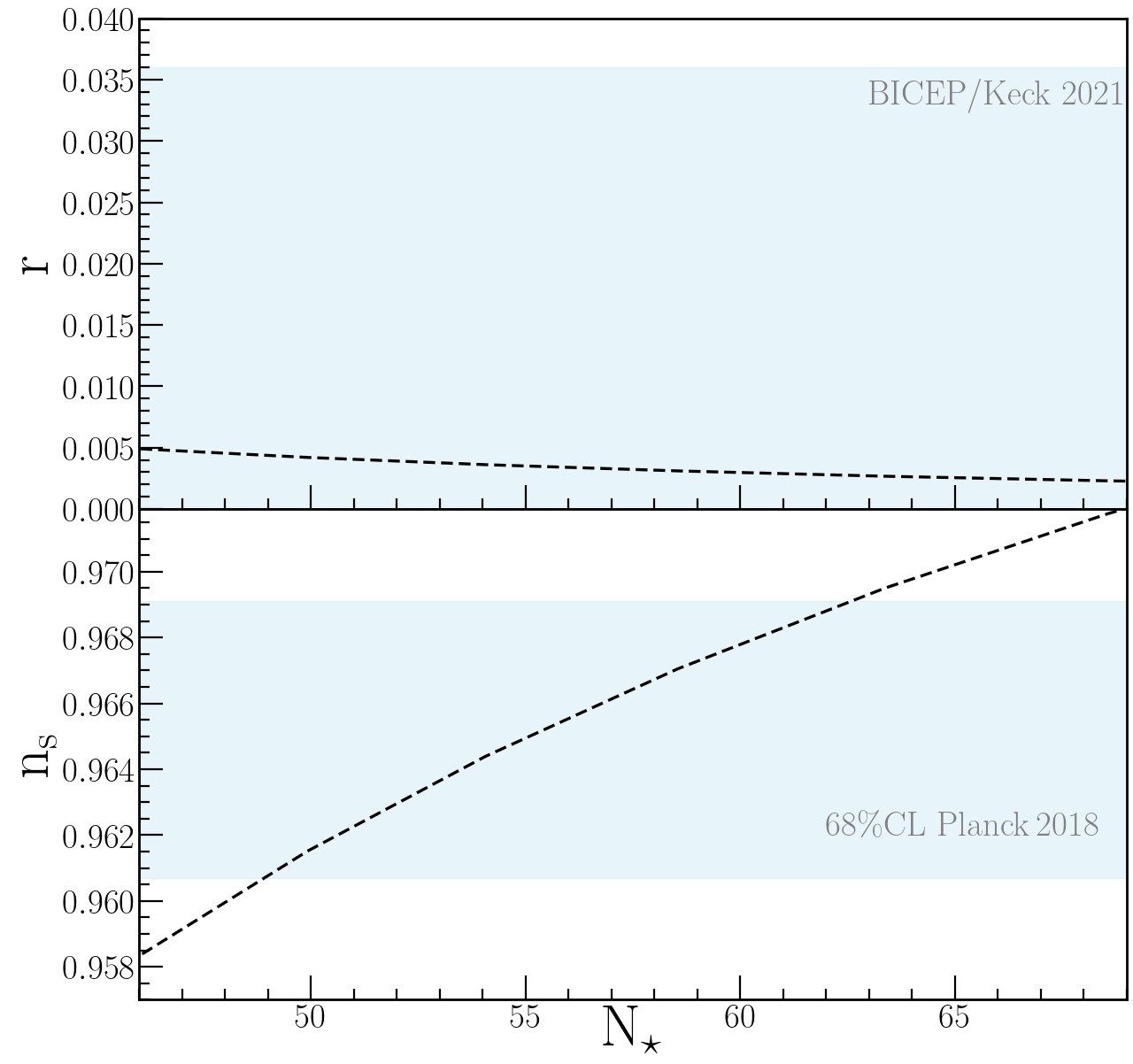}
    \caption{The spectral index $n_s$ and tensor-to-scalar ratio $r$ as a function of $N_\star$. Experimentally allowed ranges for these quantities are shown in cyan bands. }
    \label{fig:nsr}
\end{figure}
Since Eq.~\eqref{eq:Vphi} describes an approximate single-field inflationary setup, we can utilize it to analyze the evolution of the inflationary phase and associated predictions. 
We choose the relevant parameters such that the inflationary trajectory is preserved and the dynamics of $\theta_2$ negligibly affect inflation.
As discussed above, non-minimal couplings are responsible for flattening the potential in the large field limit, and the Affleck-Dine mechanism is realized via the motion of the dynamical field $\theta_2$. The amount of the asymmetry generated is determined by the size of the non-trivial motion induced in $\theta_2$ sourced by inflation. Since during inflationary phase, $m, \tilde\mu, \hat \mu \ll \phi$, and $\lambda^\prime$ is small, the quartic coupling $\widetilde \lambda$ dominates the inflation dynmaics.  Then the scalar potential Eq.~\eqref{eq:Vphi}, in this regime, can be written as 
\begin{align}
U(\chi)\simeq \frac{3}{4} m^2_s M^2_\mathrm{pl} \left( 1-e^{-\sqrt{\frac{2}{3}}\chi/M_\mathrm{pl}} \right)^2, \label{eq:UU}  
\end{align}
which mirrors the Starobinsky form~\cite{Starobinsky:1980te}, and therefore, inflationary observables are expected to be in full agreement with inflationary observations.  

From the scalar potential Eq.~\ref{eq:UU}, the slow-roll parameter is given by
\begin{align}
    \epsilon\simeq \frac{3}{4}N^{-2}_*,
\end{align}
where $N_*$ represents the  e-foldings number from the  horizon
exit to the end of the inflation, which is required to be in the range $50< N_*< 60$. $N_*$ is computed using 
\begin{align}
N_*\simeq M^{-2}_\mathrm{pl} \int^{\chi_*}_{\chi_\mathrm{end}} d\chi  \frac{V}{dV/d\chi}.
\end{align} 
Two of the observables, spectral index and the tensor-to-scalar ratio, are given by
\begin{align}
n_s\simeq 1-2N^{-1}_*, \;\;\;
r\simeq 12 N^{-2}_*, \label{eq:analytical}
\end{align}
which must satisfy the Planck observations~\cite{Planck:2018jri,BICEP:2021xfz}:
\begin{align}
&n_s = 0.9649 \pm 0.042 \;\;\;(68\% \mathrm{C.L.}) , 
\\&
r < 0.036 \;\;\;(95\% \mathrm{C.L.}).
\end{align}  
Using the approximated analytical formula given in Eq.~\eqref{eq:analytical}, one sees that within the viable range of e-folding number, namely $50< N_*< 60$, the spectrum index is predicted to be $0.96\lesssim n_s\lesssim 0.9667$, which is fully consistent with current measurements.  On the other hand, from Eq.~\eqref{eq:analytical}, the tensor-to-scalar ratio is predicted to be in the range $3.3\times 10^{-3}\lesssim r \lesssim 4.8\times 10^{-3}$. Interestingly, $r$ in this range can be probed by the  upcoming LiteBIRD
telescope~\cite{Hazumi:2019lys} and CMB-S4 experiment~\cite{Abazajian:2019eic}. By solving the corresponding equations numerically, the predicted spectral index and the tensor-to-scalar ratio as a function of the e-folding number are depicted in Fig.~\ref{fig:nsr}.

Furthermore, to match the measured scalar power spectrum amplitude at the horizon exit, we demand~\cite{Planck:2018vyg}
\begin{align}
A_s\simeq \frac{1}{12\pi^2} \frac{V^3}{M^6_P|dV/d\chi|^2} = 2.1\times 10^{-9}   
\end{align}
Fulfilling this condition requires
\begin{align}
\frac{\widetilde \lambda}{\xi^2} \sim 4.4\times 10^{-10},   
\end{align}
which restricts the size of the $\lambda$ coupling.

\subsection{Generation of Baryon Asymmetry}
The observed baryon asymmetry of the universe is measured to be~\cite{Planck:2018vyg} 
\begin{align}
\eta_N^\mathrm{obs} =\frac{n_B}{s} \simeq 8.5\times 10^{-11},   
\end{align}
where  $s=(2\pi^2/45)g_*T^3$ is the entropy density    and 
$n_B=-(28/79)n_X/s$ is the  baryon number density of the universe. As aforementioned, in our scenario, the asymmetry $n_X$ is generated through the Affleck-Dine mechanism~\cite{Affleck:1984fy}. This mechanism is based on inducing non-zero angular motion in the phase of a complex scalar field, which is charged under a global Abelian symmetry, such as $U(1)_X$ within our framework. In the early universe, this complex scalar field attains a large initial field value and begins to oscillate once the Hubble parameter drops below its mass. If the scalar potential includes an explicit $U(1)_X$ breaking term, this motion generates a net $X$ charge asymmetry. Therefore, if this symmetry is associated with the global $X=B$ or $X=L$ symmetries, a baryon asymmetry can be established before the electroweak phase transition.

\begin{figure}[!tb]
    \centering
\includegraphics[width=0.5\textwidth]{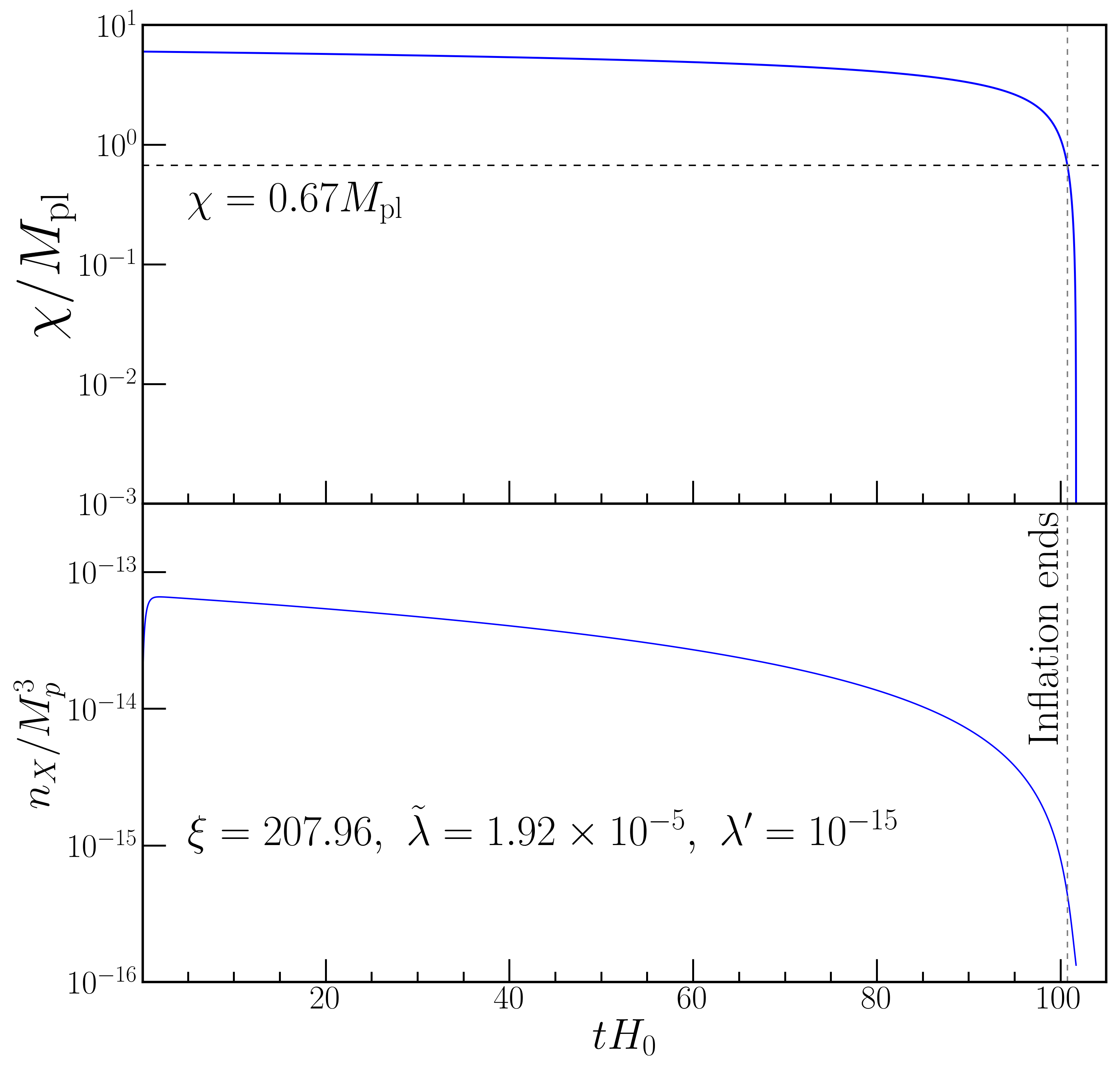}
    \caption{The evolution of $\chi$ and asymmetry $n_X$ during the inflation. We considered $\chi=6M_{\mathrm{pl}}$ and $\theta_2=0.1$ as initial conditions. The quartic couplings and non-minimal couplings to gravity are provided in Tab.~\ref{tab:BP}.}
    \label{fig:chinL}
\end{figure}

From the analysis performed above, the $U(1)_X$ charge asymmetry is generated during inflation (Fig.~\ref{fig:chinL}), and its value at the end of inflation is determined as
\begin{align}
    n_X=&Q_{B-L}\varphi_{\mathrm{end}}^2\sin^2 \alpha \sin^2 \beta \dot{\theta}_2, \\
    \approx& -Q_L\frac {4\varphi_{\mathrm{end}}^2M_{\mathrm{pl}}{\lambda^\prime}\sin \theta_\mathrm{end}}{\sqrt{3\tilde\lambda}},
\end{align}
where slow roll approximation is used. 
This net asymmetry will be transferred to the baryonic sector through
equilibrium sphaleron processes prior to the electroweak phase transition.

The field value at the end of inflation corresponds to $\chi\approx 0.67 M_\mathrm{pl}$, as can be seen from Fig.~\ref{fig:chinL}, which shows the evolution of the field $\chi$ during the inflation.  
As mentioned above, we assume that among the terms $(m, \mu, \hat\mu, \tilde \lambda, \lambda^\prime)$ in the potential Eq.~\eqref{eq:Vphi}, the quartic term with $\tilde \lambda$  dominates the inflationary dynamics, whereas the asymmetry is generated due to the presence of the  $\lambda^\prime$ coupling (which has negligible effect on the inflation).  The values of quartic coupling and non-minimal couplings to gravity that lead to this inflationary trajectory are provided in Tab.~\ref{tab:BP}.

To find the final value of the asymmetry, one needs to evolve the asymmetry after the end of inflation. The spontaneous symmetry breaking of the GUT field after the end of inflation will disrupt the inflationary trajectory described by Eq.~(\ref{eq:inflatangle}).  After the ending of the inflationary phase, the inflation starts oscillating and eventually decays. During this time, we assume that the universe goes through a matter-like epoch.  Therefore, the total baryon asymmetry generated during inflation, which is transferred to the SM thermal plasma during reheating, is given by:
\begin{align}
n_X\simeq    n_X^\mathrm{end}\times \left( \frac{a_\mathrm{end}}{a_\mathrm{rh}} \right)^{3}= n_X^\mathrm{end}\times \left( \frac{H_\mathrm{rh}}{H_\mathrm{end}} \right)^{2} .
\end{align}
For a typical parameter space, $H_\mathrm{rh}/H_\mathrm{end}\sim 10^{-2}$~\cite{Barrie:2022cub}, which leads to
\begin{equation}
    \eta_B\equiv\left. \frac {n_B}{s}  \right |_{T=T_{\mathrm{reh}}}=\eta_B^{\mathrm{obs}}\left ( \frac {n^\mathrm{end}_X   }{10^{-16} M^3_{\mathrm{pl}} } \right )\left ( \frac {g_\star}{112.75} \right )^{-\frac 14}.
\end{equation}
Here, to obtain an approximated good estimation,  following Higgs inflation, we have used the reheating temperature $T_\mathrm{rh}\simeq 2.2\times 10^{14}$ GeV~\cite{Bezrukov:2008ut,Garcia-Bellido:2008ycs} (see also, Refs.~\cite{DeCross:2015uza,Ema:2016dny,DeCross:2016fdz,DeCross:2016cbs,Sfakianakis:2018lzf,He:2018mgb,He:2020ivk,He:2020qcb}). A dedicated analysis of reheating may
be different in our scenario due to the presence of additional scalars on top of the SM Higgs, which, however, is beyond the scope of this work. 
Moreover, in our analysis, not to violate unitarity,  we consider $\xi < 350$~\cite{DeCross:2015uza,Ema:2016dny,DeCross:2016fdz}.

\textbf{Solution to the monopole problem:} Typically, GUT models suffer from the overproduction of monopoles and require inflation to dilute their density. In $SU(5)$ GUT, the adjoint Higgs breaks the GUT symmetry to the SM, and stable superheavy monopoles are formed at this stage. Typically, GUT singlets are introduced to realize inflation~\cite{Linde:1993cn,Dvali:1994ms,Pallis:2016mvm,Kaladharan:2023zbr}. Our model, on the other hand, has the special feature that the monopole problem is naturally solved since the adjoint Higgs participates in inflation and already has non-zero field values during inflation. Therefore, the associated topological defects are inflated away. The model proposed in this work is highly attractive since no additional scalar fields (such as singlets) are required beyond the minimal set for the implementation of inflation.

\textbf{Collider constraints:}
Since the VEV of the triplet is expected to be somewhat small within our scenario, its primary decay is to SM leptons. Current LHC searches of doubly-charged scalars utilizing $pp\to \Delta^{++}_1\Delta^{--}_1\to \ell^+\ell^+\ell^-\ell^-$ provides a lower limit on their masses $\gtrsim 800$ GeV~\cite{ATLAS:2017xqs}. Future colliders can probe this mass up to about 4 TeV~\cite{Du:2018eaw}. 
Therefore, neutrino mass, when combined with collider bounds and suppressing the washout  effects, leaves up with the valid range of $0.05\; \mathrm{eV} \lesssim v_\Delta \lesssim 10^{-5}$ GeV. As we will show below, a stronger bound on this VEV is obtained from the condition of preventing the generated asymmetry from being washed out.

\textbf{Lepton Flavor Violation:} In this setup, both the singly-charged ($\Delta^{\pm}_1$) and doubly-charged ($\Delta^{\pm\pm}_1$) scalars within the weak triplet lead to lepton flavor violating processes~\cite{Dinh:2012bp}. The most stringent constraint comes from the tree-level decay of $\Delta^{\pm\pm}_1$ leading to $\mu\to 3e$. On the other hand, $\mu\to e\gamma$ is generated at one-loop order via both  $\Delta^{\pm}_1$ and $\Delta^{\pm\pm}_1$ fields. Low energy experiments probing such lepton flavor violating rare decays can put a limit as high as about $10^3-10^4$ TeV on the triplet mass. The exact bounds on these masses largely depend on the size of the relevant Yukawa couplings. However, within our framework, washout constraints on the parameter space provide even stronger bounds than the lepton flavor violation.

\textbf{Washout:}
Owing to  large reheating temperature,  the triplet $\Delta_1$ will rapidly
thermalize at the beginning of the radiation era, and subsequently washout the generated asymmetry. Therefore, we require that the $LL\leftrightarrow HH$ is never in thermal equilibrium, which, when combined with the requirement of reproducing the correct neutrino mass scale, leads to $m_\Delta \lesssim 5\times 10^{11}$    GeV~\cite{Barrie:2022cub}. Moreover, to ensure that processes like $LL \leftrightarrow \Delta_1$ and  $HH \leftrightarrow \Delta_1$ do not co-exist demands $v_\Delta m^{1/2}_\Delta \lesssim 3.6\times 10^{-4}$ GeV$^{2/3}$~\cite{Barrie:2022cub}.

\begin{figure}[!tb]
    \centering
\includegraphics[width=0.5\textwidth]{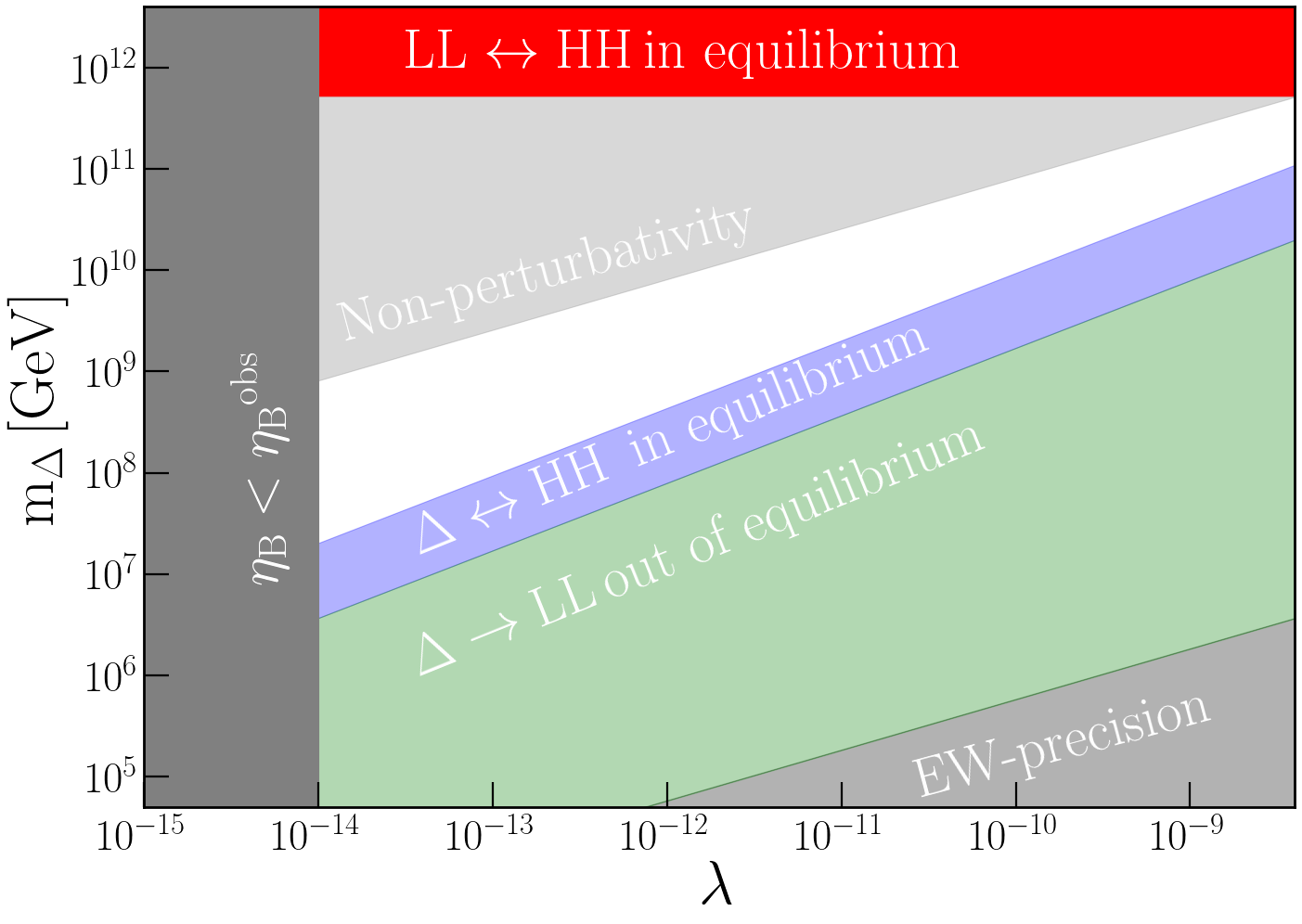}
    \caption{The allowed parameter space, in the $\lambda-m_\Delta$ plane,  is depicted in white. The vertical gray band is the regime  where generated asymmetry is not sufficient. Parameter space where the Yukawa couplings become non-perturbative is shown with a light gray regime (in the upper part of the plot). EW-precision constraint is depicted with medium gray regime (in the lower part of the plot). Constraints from the washout processes of the generated asymmetry rule out a large part of the parameter space. See text for details.  }
    \label{fig:ML}
\end{figure}

We illustrate the constraints in Figure~\ref{fig:ML}, by plotting the quartic coupling $\lambda$ versus the mass of the triplet, $m_{\Delta}$, assuming $\hat{\mu}$ is negligible and considering $\lambda^\prime = 10^{-1}\lambda$ which holds true for a typical parameter space. In this scenario, the lower bound for $\lambda \gtrsim\mathcal{O}(10^{−14})$ comes from producing the observed baryon asymmetry. The $\lambda^\prime$ term should be sufficiently small that it does not significantly alter the inflationary trajectory. Therefore, we restrict ourselves with   $\lambda \lesssim 10^{−9}$.   Moreover, as aforementioned, the approximate collider bound on the weak triplet mass, $m_\Delta \gtrsim 1$ TeV, translates into $v_\Delta \lesssim 10^{-5}$ GeV. However, we find that a more stringent constraint arises from preventing washout and generating appropriate baryon asymmetry that requires $\mu>10^{-2}$ GeV, which translates into $m_{\Delta}\gtrsim 10^{7}$ GeV and thereby providing an upper bound on the VEV $v_{\Delta}\lesssim 10^{-7}$ GeV. For a fixed value of $\lambda^\prime$, the upper limit of the $m_{\Delta}$ is provided by the perturbativity constraints of Yukawa couplings. After taking into account all these theoretical and experimental constraints, the remaining white space in Figure~\ref{fig:ML} corresponds to a viable parameter space for generating the correct baryon asymmetry of the universe.    As can be seen from Figure~\ref{fig:ML}, a viable range for the mass of the type II seesaw triplet field  is $10^{7}\;\mathrm{GeV}\lesssim m_\Delta \lesssim 10^{10}$ GeV, which is rather heavy and a very different window as obtained in Ref.~\cite{Barrie:2021mwi}. This difference is owing to the fact that unlike Ref.~\cite{Barrie:2021mwi}, in this work, we considered a renormalizable scalar potential. In our setup, the baryon asymmetry is generated by utilizing a dimension four term in the Lagrangian.

\section{Conclusions}\label{sec:con}
The origins of neutrino mass and baryon asymmetry represent two of the most profound and unresolved questions in particle physics. Simultaneously, cosmic inflation—an essential mechanism for explaining several fundamental issues in the standard Big Bang model—remains a major enigma in cosmology. In this work, we have proposed a unified theoretical framework, in particular, a minimal $SU(5)$ Grand Unified Theory, that simultaneously addresses the origins of cosmic inflation, baryon asymmetry, and neutrino mass generation. Our model integrates these phenomena into a coherent structure where inflation is governed by three essential fields: the GUT-breaking adjoint Higgs $24_H$, the Standard Model Higgs residing in the fundamental representation $5_H$, and the multiplet $15_H$, which is crucial for neutrino mass generation. Specifically, we have adopted inflation with a non-minimal coupling to gravity. The weak triplet within $15_H$, part of the inflaton, achieves a displaced vacuum value early in the universe, leading to baryon asymmetry via the Affleck-Dine mechanism.  In this setup, we have found that the viable mass range for the type II seesaw triplet field is $10^{7}\;\mathrm{GeV}\lesssim m_\Delta \lesssim 10^{10}$ GeV, which depends on the value of the quartic coupling ($\lambda_1$)--the coefficient of a renormalizable term in the scalar potential responsible for explicitly breaking a global symmetry and for the genesis of baryon asymmetry.   Additionally, the adjoint Higgs responsible for breaking the GUT symmetry acquires non-zero field values during inflation, which helps evade the monopole problem.   Within our model, the inflationary observables, especially the spectral index, show excellent agreement with experimental data, while the tensor-to-scalar ratio is expected to be probed by the future LiteBIRD and CMB-S4 experiments.  In summary, our unified approach addresses key challenges in particle physics and cosmology, providing a novel resolution to these issues.

\section*{Acknowledgement}
We thank Rabindra N. Mohapatra for discussion. The work of A.K. is supported by the U.S. Department of Energy under grant number DE-SC 0016013.  S.S. acknowledges the Center for Theoretical Underground Physics and Related Areas (CETUP* 2024) and the Institute for Underground Science at Sanford Underground Research Facility (SURF) for providing a conducive environment for the finalization of this work.  Some computing for this project was performed at the High Performance Computing Center at Oklahoma State University, supported in part through the National Science Foundation grant OAC-1531128.

\bibliographystyle{style}
\bibliography{reference}

\providecommand{\href}[2]{#2}\begingroup\raggedright\begin{thebibliography}{10}

\bibitem{Pati:1973rp}
J.~C. Pati and A.~Salam, ``{Is Baryon Number Conserved?},'' \href{http://dx.doi.org/10.1103/PhysRevLett.31.661}{{\em Phys. Rev. Lett.} {\bfseries 31} (1973) 661--664}.

\bibitem{Pati:1974yy}
J.~C. Pati and A.~Salam, ``{Lepton Number as the Fourth Color},'' \href{http://dx.doi.org/10.1103/PhysRevD.10.275}{{\em Phys. Rev. D} {\bfseries 10} (1974) 275--289}. [Erratum: Phys.Rev.D 11, 703--703 (1975)].

\bibitem{Georgi:1974sy}
H.~Georgi and S.~L. Glashow, ``{Unity of All Elementary Particle Forces},''
\href{http://dx.doi.org/10.1103/PhysRevLett.32.438}{{\em Phys. Rev. Lett.} {\bfseries 32} (1974) 438--441}.

\bibitem{Georgi:1974yf}
H.~Georgi, H.~R. Quinn, and S.~Weinberg, ``{Hierarchy of Interactions in Unified Gauge Theories},'' \href{http://dx.doi.org/10.1103/PhysRevLett.33.451}{{\em Phys. Rev. Lett.} {\bfseries 33} (1974) 451--454}.

\bibitem{Georgi:1974my}
H.~Georgi, ``{The State of the Art\textemdash{}Gauge Theories},'' \href{http://dx.doi.org/10.1063/1.2947450}{{\em AIP Conf. Proc.} {\bfseries 23} (1975) 575--582}.

\bibitem{Fritzsch:1974nn}
H.~Fritzsch and P.~Minkowski, ``{Unified Interactions of Leptons and Hadrons},'' \href{http://dx.doi.org/10.1016/0003-4916(75)90211-0}{{\em Annals Phys.} {\bfseries 93} (1975) 193--266}.

\bibitem{Minkowski:1977sc}
P.~Minkowski, ``{$\mu \to e\gamma$ at a Rate of One Out of $10^{9}$ Muon Decays?},''
\href{http://dx.doi.org/10.1016/0370-2693(77)90435-X}{{\em Phys. Lett.} {\bfseries 67B} (1977) 421--428}.

\bibitem{Yanagida:1979as}
T.~Yanagida, ``{Horizontal gauge symmetry and masses of neutrinos},''
{\em Conf. Proc.} {\bfseries C7902131} (1979) 95--99.

\bibitem{Glashow:1979nm}
S.~Glashow, ``{The Future of Elementary Particle Physics},'' \href{http://dx.doi.org/10.1007/978-1-4684-7197-7\_15}{{\em NATO Sci. Ser. B} {\bfseries 61} (1980) 687}.

\bibitem{Gell-Mann:1979vob}
M.~Gell-Mann, P.~Ramond, and R.~Slansky, ``{Complex Spinors and Unified Theories},'' {\em Conf. Proc. C} {\bfseries 790927} (1979) 315--321, \href{http://arxiv.org/abs/1306.4669}{{\ttfamily arXiv:1306.4669 [hep-th]}}.

\bibitem{Mohapatra:1979ia}
R.~N. Mohapatra and G.~Senjanovic, ``{Neutrino Mass and Spontaneous Parity Nonconservation},'' \href{http://dx.doi.org/10.1103/PhysRevLett.44.912}{{\em Phys. Rev. Lett.} {\bfseries 44} (1980) 912}.

\bibitem{Schechter:1980gr}
J.~Schechter and J.~W.~F. Valle, ``{Neutrino Masses in SU(2) x U(1) Theories},''
\href{http://dx.doi.org/10.1103/PhysRevD.22.2227}{{\em Phys. Rev.} {\bfseries D22} (1980) 2227}.

\bibitem{Schechter:1981cv}
J.~Schechter and J.~W.~F. Valle, ``{Neutrino Decay and Spontaneous Violation of Lepton Number},''
\href{http://dx.doi.org/10.1103/PhysRevD.25.774}{{\em Phys. Rev.} {\bfseries D25} (1982) 774}.

\bibitem{Dorsner:2005fq}
I.~Dorsner and P.~Fileviez~Perez, ``{Unification without supersymmetry: Neutrino mass, proton decay and light leptoquarks},'' \href{http://dx.doi.org/10.1016/j.nuclphysb.2005.06.016}{{\em Nucl. Phys. B} {\bfseries 723} (2005) 53--76}, \href{http://arxiv.org/abs/hep-ph/0504276}{{\ttfamily arXiv:hep-ph/0504276}}.

\bibitem{Dorsner:2005ii}
I.~Dorsner, P.~Fileviez~Perez, and R.~Gonzalez~Felipe, ``{Phenomenological and cosmological aspects of a minimal GUT scenario},'' \href{http://dx.doi.org/10.1016/j.nuclphysb.2006.05.006}{{\em Nucl. Phys. B} {\bfseries 747} (2006) 312--327}, \href{http://arxiv.org/abs/hep-ph/0512068}{{\ttfamily arXiv:hep-ph/0512068}}.

\bibitem{Bajc:2006ia}
B.~Bajc and G.~Senjanovic, ``{Seesaw at LHC},'' \href{http://dx.doi.org/10.1088/1126-6708/2007/08/014}{{\em JHEP} {\bfseries 08} (2007) 014},
\href{http://arxiv.org/abs/hep-ph/0612029}{{\ttfamily arXiv:hep-ph/0612029 [hep-ph]}}.

\bibitem{Dorsner:2006hw}
I.~Dorsner, P.~Fileviez~Perez, and G.~Rodrigo, ``{Fermion masses and the UV cutoff of the minimal realistic SU(5)},'' \href{http://dx.doi.org/10.1103/PhysRevD.75.125007}{{\em Phys. Rev. D} {\bfseries 75} (2007) 125007}, \href{http://arxiv.org/abs/hep-ph/0607208}{{\ttfamily arXiv:hep-ph/0607208}}.

\bibitem{Dorsner:2007fy}
I.~Dorsner and I.~Mocioiu, ``{Predictions from type II see-saw mechanism in SU(5)},'' \href{http://dx.doi.org/10.1016/j.nuclphysb.2007.12.004}{{\em Nucl. Phys.} {\bfseries B796} (2008) 123--136},
\href{http://arxiv.org/abs/0708.3332}{{\ttfamily arXiv:0708.3332 [hep-ph]}}.

\bibitem{Antusch:2021yqe}
S.~Antusch and K.~Hinze, ``{Nucleon decay in a minimal non-SUSY GUT with predicted quark-lepton Yukawa ratios},'' \href{http://dx.doi.org/10.1016/j.nuclphysb.2022.115719}{{\em Nucl. Phys. B} {\bfseries 976} (2022) 115719}, \href{http://arxiv.org/abs/2108.08080}{{\ttfamily arXiv:2108.08080 [hep-ph]}}.

\bibitem{Antusch:2022afk}
S.~Antusch, K.~Hinze, and S.~Saad, ``{Viable quark-lepton Yukawa ratios and nucleon decay predictions in SU(5) GUTs with type-II seesaw},'' \href{http://dx.doi.org/10.1016/j.nuclphysb.2022.116049}{{\em Nucl. Phys. B} {\bfseries 986} (2023) 116049}, \href{http://arxiv.org/abs/2205.01120}{{\ttfamily arXiv:2205.01120 [hep-ph]}}.

\bibitem{Calibbi:2022wko}
L.~Calibbi and X.~Gao, ``{Lepton flavor violation in minimal grand unified type II seesaw models},'' \href{http://dx.doi.org/10.1103/PhysRevD.106.095036}{{\em Phys. Rev. D} {\bfseries 106} no.~9, (2022) 095036}, \href{http://arxiv.org/abs/2206.10682}{{\ttfamily arXiv:2206.10682 [hep-ph]}}.

\bibitem{Antusch:2023kli}
S.~Antusch, K.~Hinze, and S.~Saad, ``{Quark-lepton Yukawa ratios and nucleon decay in SU(5) GUTs with type-III seesaw},'' \href{http://dx.doi.org/10.1016/j.nuclphysb.2023.116195}{{\em Nucl. Phys. B} {\bfseries 991} (2023) 116195}, \href{http://arxiv.org/abs/2301.03601}{{\ttfamily arXiv:2301.03601 [hep-ph]}}.

\bibitem{Antusch:2023mqe}
S.~Antusch, K.~Hinze, and S.~Saad, ``{Minimal SU(5) GUTs with vectorlike fermions},'' \href{http://dx.doi.org/10.1103/PhysRevD.108.095010}{{\em Phys. Rev. D} {\bfseries 108} no.~9, (2023) 095010}, \href{http://arxiv.org/abs/2308.08585}{{\ttfamily arXiv:2308.08585 [hep-ph]}}.

\bibitem{Wolfenstein:1980sf}
L.~Wolfenstein, ``{Neutrino mixing in grand unified theories},'' {\em eConf} {\bfseries C801002} (1980) 116--120.

\bibitem{Barbieri:1981yw}
R.~Barbieri, D.~V. Nanopoulos, and D.~Wyler, ``{Hierarchical Fermion Masses in SU(5)},'' \href{http://dx.doi.org/10.1016/0370-2693(81)90076-9}{{\em Phys. Lett. B} {\bfseries 103} (1981) 433--436}.

\bibitem{Perez:2016qbo}
P.~Fileviez~Perez and C.~Murgui, ``{Renormalizable SU(5) Unification},'' \href{http://dx.doi.org/10.1103/PhysRevD.94.075014}{{\em Phys. Rev.} {\bfseries D94} no.~7, (2016) 075014},
\href{http://arxiv.org/abs/1604.03377}{{\ttfamily arXiv:1604.03377 [hep-ph]}}.

\bibitem{Kumericki:2017sfc}
K.~Kumericki, T.~Mede, and I.~Picek, ``{Renormalizable SU(5) Completions of a Zee-type Neutrino Mass Model},'' \href{http://dx.doi.org/10.1103/PhysRevD.97.055012}{{\em Phys. Rev.} {\bfseries D97} no.~5, (2018) 055012},
\href{http://arxiv.org/abs/1712.05246}{{\ttfamily arXiv:1712.05246 [hep-ph]}}.

\bibitem{Saad:2019vjo}
S.~Saad, ``{Origin of a two-loop neutrino mass from SU(5) grand unification},'' \href{http://dx.doi.org/10.1103/PhysRevD.99.115016}{{\em Phys. Rev.} {\bfseries D99} no.~11, (2019) 115016},
\href{http://arxiv.org/abs/1902.11254}{{\ttfamily arXiv:1902.11254 [hep-ph]}}.

\bibitem{Dorsner:2019vgf}
I.~Dor\v{s}ner and S.~Saad, ``{Towards Minimal $SU(5)$},'' \href{http://dx.doi.org/10.1103/PhysRevD.101.015009}{{\em Phys. Rev. D} {\bfseries 101} no.~1, (2020) 015009}, \href{http://arxiv.org/abs/1910.09008}{{\ttfamily arXiv:1910.09008 [hep-ph]}}.

\bibitem{Dorsner:2021qwg}
I.~Dor\v{s}ner, E.~D\v{z}aferovi\'c-Ma\v{s}i\'c, and S.~Saad, ``{Parameter space exploration of the minimal SU(5) unification},'' \href{http://dx.doi.org/10.1103/PhysRevD.104.015023}{{\em Phys. Rev. D} {\bfseries 104} no.~1, (2021) 015023}, \href{http://arxiv.org/abs/2105.01678}{{\ttfamily arXiv:2105.01678 [hep-ph]}}.

\bibitem{Antusch:2023jok}
S.~Antusch, I.~Dor\v{s}ner, K.~Hinze, and S.~Saad, ``{Fully testable axion dark matter within a minimal SU(5) GUT},'' \href{http://dx.doi.org/10.1103/PhysRevD.108.015025}{{\em Phys. Rev. D} {\bfseries 108} no.~1, (2023) 015025}, \href{http://arxiv.org/abs/2301.00809}{{\ttfamily arXiv:2301.00809 [hep-ph]}}.

\bibitem{Dorsner:2024jiy}
I.~Dor\v{s}ner, E.~D\v{z}aferovi\'c-Ma\v{s}i\'c, S.~Fajfer, and S.~Saad, ``{Gauge and scalar boson mediated proton decay in a predictive SU(5) GUT model},'' \href{http://dx.doi.org/10.1103/PhysRevD.109.075023}{{\em Phys. Rev. D} {\bfseries 109} no.~7, (2024) 075023}, \href{http://arxiv.org/abs/2401.16907}{{\ttfamily arXiv:2401.16907 [hep-ph]}}.

\bibitem{Klein:2019jgb}
C.~Klein, M.~Lindner, and S.~Vogl, ``{Radiative neutrino masses and successful $SU(5)$ unification},'' \href{http://dx.doi.org/10.1103/PhysRevD.100.075024}{{\em Phys. Rev.} {\bfseries D100} no.~7, (2019) 075024},
\href{http://arxiv.org/abs/1907.05328}{{\ttfamily arXiv:1907.05328 [hep-ph]}}.

\bibitem{Barrie:2021mwi}
N.~D. Barrie, C.~Han, and H.~Murayama, ``{Affleck-Dine Leptogenesis from Higgs Inflation},'' \href{http://dx.doi.org/10.1103/PhysRevLett.128.141801}{{\em Phys. Rev. Lett.} {\bfseries 128} no.~14, (2022) 141801}, \href{http://arxiv.org/abs/2106.03381}{{\ttfamily arXiv:2106.03381 [hep-ph]}}.

\bibitem{Affleck:1984fy}
I.~Affleck and M.~Dine, ``{A New Mechanism for Baryogenesis},'' \href{http://dx.doi.org/10.1016/0550-3213(85)90021-5}{{\em Nucl. Phys. B} {\bfseries 249} (1985) 361--380}.

\bibitem{Guth:1980zm}
A.~H. Guth, ``{The Inflationary Universe: A Possible Solution to the Horizon and Flatness Problems},'' \href{http://dx.doi.org/10.1103/PhysRevD.23.347}{{\em Phys. Rev. D} {\bfseries 23} (1981) 347--356}.

\bibitem{Albrecht:1982wi}
A.~Albrecht and P.~J. Steinhardt, ``{Cosmology for Grand Unified Theories with Radiatively Induced Symmetry Breaking},'' \href{http://dx.doi.org/10.1103/PhysRevLett.48.1220}{{\em Phys. Rev. Lett.} {\bfseries 48} (1982) 1220--1223}.

\bibitem{Linde:1981mu}
A.~D. Linde, ``{A New Inflationary Universe Scenario: A Possible Solution of the Horizon, Flatness, Homogeneity, Isotropy and Primordial Monopole Problems},'' \href{http://dx.doi.org/10.1016/0370-2693(82)91219-9}{{\em Phys. Lett.} {\bfseries 108B} (1982) 389--393}.
[Adv. Ser. Astrophys. Cosmol.3,149(1987)].

\bibitem{Linde:1983gd}
A.~D. Linde, ``{Chaotic Inflation},'' \href{http://dx.doi.org/10.1016/0370-2693(83)90837-7}{{\em Phys. Lett. B} {\bfseries 129} (1983) 177--181}.

\bibitem{Barrie:2022cub}
N.~D. Barrie, C.~Han, and H.~Murayama, ``{Type II Seesaw leptogenesis},'' \href{http://dx.doi.org/10.1007/JHEP05(2022)160}{{\em JHEP} {\bfseries 05} (2022) 160}, \href{http://arxiv.org/abs/2204.08202}{{\ttfamily arXiv:2204.08202 [hep-ph]}}.

\bibitem{Brout:1977ix}
R.~Brout, F.~Englert, and E.~Gunzig, ``{The Creation of the Universe as a Quantum Phenomenon},'' \href{http://dx.doi.org/10.1016/0003-4916(78)90176-8}{{\em Annals Phys.} {\bfseries 115} (1978) 78}.

\bibitem{Sato:1981qmu}
K.~Sato, ``{First-order phase transition of a vacuum and the expansion of the Universe},'' \href{http://dx.doi.org/10.1093/mnras/195.3.467}{{\em Mon. Not. Roy. Astron. Soc.} {\bfseries 195} no.~3, (1981) 467--479}.

\bibitem{Albrecht:1982mp}
A.~Albrecht, P.~J. Steinhardt, M.~S. Turner, and F.~Wilczek, ``{Reheating an Inflationary Universe},''
\href{http://dx.doi.org/10.1103/PhysRevLett.48.1437}{{\em Phys. Rev. Lett.} {\bfseries 48} (1982) 1437}.

\bibitem{Kibble:1976sj}
T.~W.~B. Kibble, ``{Topology of Cosmic Domains and Strings},'' \href{http://dx.doi.org/10.1088/0305-4470/9/8/029}{{\em J. Phys. A} {\bfseries 9} (1976) 1387--1398}.

\bibitem{Preskill:1979zi}
J.~Preskill, ``{Cosmological Production of Superheavy Magnetic Monopoles},'' \href{http://dx.doi.org/10.1103/PhysRevLett.43.1365}{{\em Phys. Rev. Lett.} {\bfseries 43} (1979) 1365}.

\bibitem{Leontaris:2016jty}
G.~K. Leontaris, N.~Okada, and Q.~Shafi, ``{Non-minimal quartic inflation in supersymmetric $SO(10)$},'' \href{http://dx.doi.org/10.1016/j.physletb.2016.12.038}{{\em Phys. Lett. B} {\bfseries 765} (2017) 256--259}, \href{http://arxiv.org/abs/1611.10196}{{\ttfamily arXiv:1611.10196 [hep-ph]}}.

\bibitem{Mohapatra:2021aig}
R.~N. Mohapatra and N.~Okada, ``{Affleck-Dine baryogenesis with observable neutron-antineutron oscillation},'' \href{http://dx.doi.org/10.1103/PhysRevD.104.055030}{{\em Phys. Rev. D} {\bfseries 104} no.~5, (2021) 055030}, \href{http://arxiv.org/abs/2107.01514}{{\ttfamily arXiv:2107.01514 [hep-ph]}}.

\bibitem{Mohapatra:2022ngo}
R.~N. Mohapatra and N.~Okada, ``{Neutrino mass from Affleck-Dine leptogenesis and WIMP dark matter},'' \href{http://dx.doi.org/10.1007/JHEP03(2022)092}{{\em JHEP} {\bfseries 03} (2022) 092}, \href{http://arxiv.org/abs/2201.06151}{{\ttfamily arXiv:2201.06151 [hep-ph]}}.

\bibitem{Langacker:1980js}
P.~Langacker, ``{Grand Unified Theories and Proton Decay},'' \href{http://dx.doi.org/10.1016/0370-1573(81)90059-4}{{\em Phys. Rept.} {\bfseries 72} (1981) 185}.

\bibitem{Dev:2022jbf}
P.~S.~B. Dev {\em et~al.}, ``{Searches for baryon number violation in neutrino experiments: a white paper},'' \href{http://dx.doi.org/10.1088/1361-6471/ad1658}{{\em J. Phys. G} {\bfseries 51} no.~3, (2024) 033001}, \href{http://arxiv.org/abs/2203.08771}{{\ttfamily arXiv:2203.08771 [hep-ex]}}.

\bibitem{Super-Kamiokande:2020wjk}
{\bfseries Super-Kamiokande} Collaboration, A.~Takenaka {\em et~al.}, ``{Search for proton decay via $p\to e^+\pi^0$ and $p\to \mu^+\pi^0$ with an enlarged fiducial volume in Super-Kamiokande I-IV},'' \href{http://dx.doi.org/10.1103/PhysRevD.102.112011}{{\em Phys. Rev. D} {\bfseries 102} no.~11, (2020) 112011}, \href{http://arxiv.org/abs/2010.16098}{{\ttfamily arXiv:2010.16098 [hep-ex]}}.

\bibitem{Magg:1980ut}
M.~Magg and C.~Wetterich, ``{Neutrino Mass Problem and Gauge Hierarchy},'' \href{http://dx.doi.org/10.1016/0370-2693(80)90825-4}{{\em Phys. Lett. B} {\bfseries 94} (1980) 61--64}.

\bibitem{Lazarides:1980nt}
G.~Lazarides, Q.~Shafi, and C.~Wetterich, ``{Proton Lifetime and Fermion Masses in an SO(10) Model},'' \href{http://dx.doi.org/10.1016/0550-3213(81)90354-0}{{\em Nucl. Phys. B} {\bfseries 181} (1981) 287--300}.

\bibitem{Mohapatra:1980yp}
R.~N. Mohapatra and G.~Senjanovic, ``{Neutrino Masses and Mixings in Gauge Models with Spontaneous Parity Violation},'' \href{http://dx.doi.org/10.1103/PhysRevD.23.165}{{\em Phys. Rev. D} {\bfseries 23} (1981) 165}.

\bibitem{Wilczek:1979hc}
F.~Wilczek and A.~Zee, ``{Operator Analysis of Nucleon Decay},'' \href{http://dx.doi.org/10.1103/PhysRevLett.43.1571}{{\em Phys. Rev. Lett.} {\bfseries 43} (1979) 1571--1573}.

\bibitem{Barrie:2024yhj}
N.~D. Barrie and C.~Han, ``{Affleck-Dine Dirac Leptogenesis},'' \href{http://arxiv.org/abs/2402.15245}{{\ttfamily arXiv:2402.15245 [hep-ph]}}.

\bibitem{Chernikov:1968zm}
N.~A. Chernikov and E.~A. Tagirov, ``{Quantum theory of scalar fields in de Sitter space-time},'' {\em Ann. Inst. H. Poincare A Phys. Theor.} {\bfseries 9} (1968) 109.

\bibitem{Planck:2018jri}
{\bfseries Planck} Collaboration, Y.~Akrami {\em et~al.}, ``{Planck 2018 results. X. Constraints on inflation},'' \href{http://dx.doi.org/10.1051/0004-6361/201833887}{{\em Astron. Astrophys.} {\bfseries 641} (2020) A10}, \href{http://arxiv.org/abs/1807.06211}{{\ttfamily arXiv:1807.06211 [astro-ph.CO]}}.

\bibitem{Bezrukov:2007ep}
F.~L. Bezrukov and M.~Shaposhnikov, ``{The Standard Model Higgs boson as the inflaton},'' \href{http://dx.doi.org/10.1016/j.physletb.2007.11.072}{{\em Phys. Lett. B} {\bfseries 659} (2008) 703--706}, \href{http://arxiv.org/abs/0710.3755}{{\ttfamily arXiv:0710.3755 [hep-th]}}.

\bibitem{Bezrukov:2008ut}
F.~Bezrukov, D.~Gorbunov, and M.~Shaposhnikov, ``{On initial conditions for the Hot Big Bang},'' \href{http://dx.doi.org/10.1088/1475-7516/2009/06/029}{{\em JCAP} {\bfseries 06} (2009) 029}, \href{http://arxiv.org/abs/0812.3622}{{\ttfamily arXiv:0812.3622 [hep-ph]}}.

\bibitem{Garcia-Bellido:2008ycs}
J.~Garcia-Bellido, D.~G. Figueroa, and J.~Rubio, ``{Preheating in the Standard Model with the Higgs-Inflaton coupled to gravity},'' \href{http://dx.doi.org/10.1103/PhysRevD.79.063531}{{\em Phys. Rev. D} {\bfseries 79} (2009) 063531}, \href{http://arxiv.org/abs/0812.4624}{{\ttfamily arXiv:0812.4624 [hep-ph]}}.

\bibitem{Barbon:2009ya}
J.~L.~F. Barbon and J.~R. Espinosa, ``{On the Naturalness of Higgs Inflation},'' \href{http://dx.doi.org/10.1103/PhysRevD.79.081302}{{\em Phys. Rev. D} {\bfseries 79} (2009) 081302}, \href{http://arxiv.org/abs/0903.0355}{{\ttfamily arXiv:0903.0355 [hep-ph]}}.

\bibitem{Barvinsky:2009fy}
A.~O. Barvinsky, A.~Y. Kamenshchik, C.~Kiefer, A.~A. Starobinsky, and C.~Steinwachs, ``{Asymptotic freedom in inflationary cosmology with a non-minimally coupled Higgs field},'' \href{http://dx.doi.org/10.1088/1475-7516/2009/12/003}{{\em JCAP} {\bfseries 12} (2009) 003}, \href{http://arxiv.org/abs/0904.1698}{{\ttfamily arXiv:0904.1698 [hep-ph]}}.

\bibitem{Bezrukov:2009db}
F.~Bezrukov and M.~Shaposhnikov, ``{Standard Model Higgs boson mass from inflation: Two loop analysis},'' \href{http://dx.doi.org/10.1088/1126-6708/2009/07/089}{{\em JHEP} {\bfseries 07} (2009) 089}, \href{http://arxiv.org/abs/0904.1537}{{\ttfamily arXiv:0904.1537 [hep-ph]}}.

\bibitem{Giudice:2010ka}
G.~F. Giudice and H.~M. Lee, ``{Unitarizing Higgs Inflation},'' \href{http://dx.doi.org/10.1016/j.physletb.2010.10.035}{{\em Phys. Lett. B} {\bfseries 694} (2011) 294--300}, \href{http://arxiv.org/abs/1010.1417}{{\ttfamily arXiv:1010.1417 [hep-ph]}}.

\bibitem{Bezrukov:2010jz}
F.~Bezrukov, A.~Magnin, M.~Shaposhnikov, and S.~Sibiryakov, ``{Higgs inflation: consistency and generalisations},'' \href{http://dx.doi.org/10.1007/JHEP01(2011)016}{{\em JHEP} {\bfseries 01} (2011) 016}, \href{http://arxiv.org/abs/1008.5157}{{\ttfamily arXiv:1008.5157 [hep-ph]}}.

\bibitem{Burgess:2010zq}
C.~P. Burgess, H.~M. Lee, and M.~Trott, ``{Comment on Higgs Inflation and Naturalness},'' \href{http://dx.doi.org/10.1007/JHEP07(2010)007}{{\em JHEP} {\bfseries 07} (2010) 007}, \href{http://arxiv.org/abs/1002.2730}{{\ttfamily arXiv:1002.2730 [hep-ph]}}.

\bibitem{Lebedev:2011aq}
O.~Lebedev and H.~M. Lee, ``{Higgs Portal Inflation},'' \href{http://dx.doi.org/10.1140/epjc/s10052-011-1821-0}{{\em Eur. Phys. J. C} {\bfseries 71} (2011) 1821}, \href{http://arxiv.org/abs/1105.2284}{{\ttfamily arXiv:1105.2284 [hep-ph]}}.

\bibitem{Lee:2018esk}
H.~M. Lee, ``{Light inflaton completing Higgs inflation},'' \href{http://dx.doi.org/10.1103/PhysRevD.98.015020}{{\em Phys. Rev. D} {\bfseries 98} no.~1, (2018) 015020}, \href{http://arxiv.org/abs/1802.06174}{{\ttfamily arXiv:1802.06174 [hep-ph]}}.

\bibitem{Choi:2019osi}
S.-M. Choi, Y.-J. Kang, H.~M. Lee, and K.~Yamashita, ``{Unitary inflaton as decaying dark matter},'' \href{http://dx.doi.org/10.1007/JHEP05(2019)060}{{\em JHEP} {\bfseries 05} (2019) 060}, \href{http://arxiv.org/abs/1902.03781}{{\ttfamily arXiv:1902.03781 [hep-ph]}}.

\bibitem{Sopov:2022bog}
A.~H. Sopov and R.~R. Volkas, ``{VISH\ensuremath{\nu}: solving five Standard Model shortcomings with a Poincar\'e-protected electroweak scale},'' \href{http://dx.doi.org/10.1016/j.dark.2023.101381}{{\em Phys. Dark Univ.} {\bfseries 42} (2023) 101381}, \href{http://arxiv.org/abs/2206.11598}{{\ttfamily arXiv:2206.11598 [hep-ph]}}.

\bibitem{Kaiser:2010ps}
D.~I. Kaiser, ``{Conformal Transformations with Multiple Scalar Fields},'' \href{http://dx.doi.org/10.1103/PhysRevD.81.084044}{{\em Phys. Rev. D} {\bfseries 81} (2010) 084044}, \href{http://arxiv.org/abs/1003.1159}{{\ttfamily arXiv:1003.1159 [gr-qc]}}.

\bibitem{Starobinsky:1980te}
A.~A. Starobinsky, ``{A New Type of Isotropic Cosmological Models Without Singularity},'' \href{http://dx.doi.org/10.1016/0370-2693(80)90670-X}{{\em Phys. Lett. B} {\bfseries 91} (1980) 99--102}.

\bibitem{BICEP:2021xfz}
{\bfseries BICEP, Keck} Collaboration, P.~A.~R. Ade {\em et~al.}, ``{Improved Constraints on Primordial Gravitational Waves using Planck, WMAP, and BICEP/Keck Observations through the 2018 Observing Season},'' \href{http://dx.doi.org/10.1103/PhysRevLett.127.151301}{{\em Phys. Rev. Lett.} {\bfseries 127} no.~15, (2021) 151301}, \href{http://arxiv.org/abs/2110.00483}{{\ttfamily arXiv:2110.00483 [astro-ph.CO]}}.

\bibitem{Hazumi:2019lys}
M.~Hazumi {\em et~al.}, ``{LiteBIRD: A Satellite for the Studies of B-Mode Polarization and Inflation from Cosmic Background Radiation Detection},'' \href{http://dx.doi.org/10.1007/s10909-019-02150-5}{{\em J. Low Temp. Phys.} {\bfseries 194} no.~5-6, (2019) 443--452}.

\bibitem{Abazajian:2019eic}
K.~Abazajian {\em et~al.}, ``{CMB-S4 Science Case, Reference Design, and Project Plan},'' \href{http://arxiv.org/abs/1907.04473}{{\ttfamily arXiv:1907.04473 [astro-ph.IM]}}.

\bibitem{Planck:2018vyg}
{\bfseries Planck} Collaboration, N.~Aghanim {\em et~al.}, ``{Planck 2018 results. VI. Cosmological parameters},'' \href{http://dx.doi.org/10.1051/0004-6361/201833910}{{\em Astron. Astrophys.} {\bfseries 641} (2020) A6}, \href{http://arxiv.org/abs/1807.06209}{{\ttfamily arXiv:1807.06209 [astro-ph.CO]}}. [Erratum: Astron.Astrophys. 652, C4 (2021)].

\bibitem{DeCross:2015uza}
M.~P. DeCross, D.~I. Kaiser, A.~Prabhu, C.~Prescod-Weinstein, and E.~I. Sfakianakis, ``{Preheating after Multifield Inflation with Nonminimal Couplings, I: Covariant Formalism and Attractor Behavior},'' \href{http://dx.doi.org/10.1103/PhysRevD.97.023526}{{\em Phys. Rev. D} {\bfseries 97} no.~2, (2018) 023526}, \href{http://arxiv.org/abs/1510.08553}{{\ttfamily arXiv:1510.08553 [astro-ph.CO]}}.

\bibitem{Ema:2016dny}
Y.~Ema, R.~Jinno, K.~Mukaida, and K.~Nakayama, ``{Violent Preheating in Inflation with Nonminimal Coupling},'' \href{http://dx.doi.org/10.1088/1475-7516/2017/02/045}{{\em JCAP} {\bfseries 02} (2017) 045}, \href{http://arxiv.org/abs/1609.05209}{{\ttfamily arXiv:1609.05209 [hep-ph]}}.

\bibitem{DeCross:2016fdz}
M.~P. DeCross, D.~I. Kaiser, A.~Prabhu, C.~Prescod-Weinstein, and E.~I. Sfakianakis, ``{Preheating after multifield inflation with nonminimal couplings, II: Resonance Structure},'' \href{http://dx.doi.org/10.1103/PhysRevD.97.023527}{{\em Phys. Rev. D} {\bfseries 97} no.~2, (2018) 023527}, \href{http://arxiv.org/abs/1610.08868}{{\ttfamily arXiv:1610.08868 [astro-ph.CO]}}.

\bibitem{DeCross:2016cbs}
M.~P. DeCross, D.~I. Kaiser, A.~Prabhu, C.~Prescod-Weinstein, and E.~I. Sfakianakis, ``{Preheating after multifield inflation with nonminimal couplings, III: Dynamical spacetime results},'' \href{http://dx.doi.org/10.1103/PhysRevD.97.023528}{{\em Phys. Rev. D} {\bfseries 97} no.~2, (2018) 023528}, \href{http://arxiv.org/abs/1610.08916}{{\ttfamily arXiv:1610.08916 [astro-ph.CO]}}.

\bibitem{Sfakianakis:2018lzf}
E.~I. Sfakianakis and J.~van~de Vis, ``{Preheating after Higgs Inflation: Self-Resonance and Gauge boson production},'' \href{http://dx.doi.org/10.1103/PhysRevD.99.083519}{{\em Phys. Rev. D} {\bfseries 99} no.~8, (2019) 083519}, \href{http://arxiv.org/abs/1810.01304}{{\ttfamily arXiv:1810.01304 [hep-ph]}}.

\bibitem{He:2018mgb}
M.~He, R.~Jinno, K.~Kamada, S.~C. Park, A.~A. Starobinsky, and J.~Yokoyama, ``{On the violent preheating in the mixed Higgs-$R^2$ inflationary model},'' \href{http://dx.doi.org/10.1016/j.physletb.2019.02.008}{{\em Phys. Lett. B} {\bfseries 791} (2019) 36--42}, \href{http://arxiv.org/abs/1812.10099}{{\ttfamily arXiv:1812.10099 [hep-ph]}}.

\bibitem{He:2020ivk}
M.~He, R.~Jinno, K.~Kamada, A.~A. Starobinsky, and J.~Yokoyama, ``{Occurrence of tachyonic preheating in the mixed Higgs-R$^2$ model},'' \href{http://dx.doi.org/10.1088/1475-7516/2021/01/066}{{\em JCAP} {\bfseries 01} (2021) 066}, \href{http://arxiv.org/abs/2007.10369}{{\ttfamily arXiv:2007.10369 [hep-ph]}}.

\bibitem{He:2020qcb}
M.~He, ``{Perturbative Reheating in the Mixed Higgs-$R^2$ Model},'' \href{http://dx.doi.org/10.1088/1475-7516/2021/05/021}{{\em JCAP} {\bfseries 05} (2021) 021}, \href{http://arxiv.org/abs/2010.11717}{{\ttfamily arXiv:2010.11717 [hep-ph]}}.

\bibitem{Linde:1993cn}
A.~D. Linde, ``{Hybrid inflation},'' \href{http://dx.doi.org/10.1103/PhysRevD.49.748}{{\em Phys. Rev. D} {\bfseries 49} (1994) 748--754}, \href{http://arxiv.org/abs/astro-ph/9307002}{{\ttfamily arXiv:astro-ph/9307002}}.

\bibitem{Dvali:1994ms}
G.~R. Dvali, Q.~Shafi, and R.~K. Schaefer, ``{Large scale structure and supersymmetric inflation without fine tuning},'' \href{http://dx.doi.org/10.1103/PhysRevLett.73.1886}{{\em Phys. Rev. Lett.} {\bfseries 73} (1994) 1886--1889}, \href{http://arxiv.org/abs/hep-ph/9406319}{{\ttfamily arXiv:hep-ph/9406319}}.

\bibitem{Pallis:2016mvm}
C.~Pallis and N.~Toumbas, ``{Starobinsky Inflation: From Non-SUSY To SUGRA Realizations},'' \href{http://dx.doi.org/10.1155/2017/6759267}{{\em Adv. High Energy Phys.} {\bfseries 2017} (2017) 6759267}, \href{http://arxiv.org/abs/1612.09202}{{\ttfamily arXiv:1612.09202 [hep-ph]}}.

\bibitem{Kaladharan:2023zbr}
A.~Kaladharan and S.~Saad, ``{Fermion mass, axion dark matter, and leptogenesis in SO(10) GUT},'' \href{http://dx.doi.org/10.1103/PhysRevD.109.055010}{{\em Phys. Rev. D} {\bfseries 109} no.~5, (2024) 055010}, \href{http://arxiv.org/abs/2308.04497}{{\ttfamily arXiv:2308.04497 [hep-ph]}}.

\bibitem{ATLAS:2017xqs}
{\bfseries ATLAS} Collaboration, M.~Aaboud {\em et~al.}, ``{Search for doubly charged Higgs boson production in multi-lepton final states with the ATLAS detector using proton\textendash{}proton collisions at $\sqrt{s}=13\,\text {TeV}$},'' \href{http://dx.doi.org/10.1140/epjc/s10052-018-5661-z}{{\em Eur. Phys. J. C} {\bfseries 78} no.~3, (2018) 199}, \href{http://arxiv.org/abs/1710.09748}{{\ttfamily arXiv:1710.09748 [hep-ex]}}.

\bibitem{Du:2018eaw}
Y.~Du, A.~Dunbrack, M.~J. Ramsey-Musolf, and J.-H. Yu, ``{Type-II Seesaw Scalar Triplet Model at a 100 TeV $pp$ Collider: Discovery and Higgs Portal Coupling Determination},'' \href{http://dx.doi.org/10.1007/JHEP01(2019)101}{{\em JHEP} {\bfseries 01} (2019) 101}, \href{http://arxiv.org/abs/1810.09450}{{\ttfamily arXiv:1810.09450 [hep-ph]}}.

\bibitem{Dinh:2012bp}
D.~N. Dinh, A.~Ibarra, E.~Molinaro, and S.~T. Petcov, ``{The $\mu - e$ Conversion in Nuclei, $\mu \to e \gamma, \mu \to 3e$ Decays and TeV Scale See-Saw Scenarios of Neutrino Mass Generation},'' \href{http://dx.doi.org/10.1007/JHEP08(2012)125}{{\em JHEP} {\bfseries 08} (2012) 125}, \href{http://arxiv.org/abs/1205.4671}{{\ttfamily arXiv:1205.4671 [hep-ph]}}. [Erratum: JHEP 09, 023 (2013)].

\end{thebibliography}\endgroup
\end{document}